# Self-referencing photothermal common-path interferometry to measure absorption of $Si_3N_4$ membranes for laser-light sails

Tanuj Kumar[1‡], Demeng Feng[1‡], Shenwei Yin[1], Merlin Mah[2], Phyo Lin[2], Margaret A. Fortman[3], Gabriel R. Jaffe[3], Chenghao Wan[1,4], Hongyan Mei[1], Yuzhe Xiao[1,5], Ron Synowicki[6], Ronald J. Warzoha[7], Victor W. Brar[3], Joseph J. Talghader[2], Mikhail A. Kats[1♦]

[1]Department of Electrical and Computer Engineering, University of Wisconsin–Madison, Madison, WI 53706, USA
[2]Department of Electrical and Computer Engineering, University of Minnesota–Twin Cities, MN 55455, USA
[3]Department of Physics, University of Wisconsin–Madison, Madison, WI 53706, USA
[4]Department of Electrical Engineering, Stanford University, Stanford, CA 94305, USA
[5]Department of Physics, University of North Texas, Denton, TX, 76203, USA
[6]J. A. Woollam Co. Inc., 645 M St Suite 102, Lincoln, NE 68508, USA
[7]Department of Mechanical Engineering, United States Naval Academy, Annapolis, MD 21402, USA

## Abstract

Laser-light sails are a spacecraft concept wherein lightweight "sails" are propelled by high-intensity lasers. We investigated the near-infrared absorption of free-standing membranes of stoichiometric silicon nitride ($Si_3N_4$), a candidate sail material. To resolve the small but non-zero optical loss, we used photothermal common-path interferometry (PCI), for which we developed a self-referencing modality where a PCI measurement is performed twice: once on a bare membrane, and a second time with monolayer graphene deposited on the membrane. The graphene increases the absorption of the sample by orders of magnitude, such that it can be measured by ellipsometry, without significantly affecting the thermal properties. We measured the absorption coefficient of $Si_3N_4$ to be (1.5-3) $\times 10^{-2}$ $cm^{-1}$ at 1064 nm, making it a suitable sail material for laser intensities as high as ~10 $GW/m^2$. By comparison, silicon-rich "low stress" $SiN_x$ ($x$~1), with a measured absorption coefficient of approximately 8 $cm^{-1}$, is unlikely to survive such high laser intensities. Our self-referencing technique enables testing of low-loss membranes of various materials for laser sails and other applications.

‡ - equal contribution
♦ - corresponding author. Email: mkats@wisc.edu

## Introduction

Precise measurement of optical absorption in low-loss materials is important for applications from on-chip photonics to sensitive experiments like gravitational-wave detection in LIGO[1–4]. An application of recent interest is the development of light sails propelled by high-power lasers from Earth, where laser intensities as high as 10-100 GW/m$^2$ are being considered[5–7].

The choice of materials for laser sails is important to achieve efficient acceleration and maintain sail integrity under intense illumination, with requirements that include low linear and nonlinear absorption, high refractive index (to maximize reflectivity with the smallest amount of material[8,9]), low mass density, and high thermal conductivity[9–11]. Stoichiometric silicon nitride ($Si_3N_4$) is being considered for laser-sail applications due to its moderately high refractive index (~2) and low loss in the near-infrared, large bandgap of at least ~3.3 eV[12,13] that is inconducive to near-infrared two-photon absorption, and high extinction coefficient in the mid-infrared which can aid in radiative cooling[8].

Measuring precise values of the absorption coefficients of low-loss materials, such as the absorption coefficient of $Si_3N_4$ in the near infrared, is challenging for the same reason that it is useful (i.e., because the losses are low), and conventional techniques such as ellipsometry and reflection/transmission spectroscopy can be insufficient. This is especially the case for samples with a membrane form factor, such as those required for light sails. There have been several measurements of $Si_3N_4$ using cavity ring-down spectroscopy with microfabricated $Si_3N_4$ waveguide resonators[1,14,15], but it is not clear that these measurements are directly applicable for suspended membranes in free space due to the potential presence of scattering losses and other interface effects that are difficult to distinguish from absorption losses, as well as due to potential differences in material strain. There is therefore a need for direct measurement of the optical absorption of membranes of $Si_3N_4$ and other low-loss materials that could comprise suspended structures, such as layered van der Waals materials that are also being considered for laser-light sails[16–18].

In this paper, we explore photo-thermal common-path interferometry (PCI)[2,19–22] to directly measure the optical absorption of suspended low-loss membranes. In PCI, a chopped continuous-wave pump laser is incident on the material being tested, resulting in heating; the small increase in temperature results in a change of refractive index via the thermo-optic effect, and this change is measured using a probe laser at a different wavelength and incident angle compared to the pump laser[2,3,22–24]. The conversion from a PCI measurement to an absolute absorption value is not trivial, because it is a function of both optical and thermal processes, and we found that most methods in the literature[2,20–26] are difficult to use for free-standing structures (such as membranes) that have nontrivial thermal conduction to the supporting frame.

We present a new self-referencing PCI modality, in which a PCI measurement is performed on a suspended membrane of interest, and then on an identical sample onto which we have transferred a monolayer of graphene. Monolayer graphene has a well-known and large optical absorption (~2.3% in free space) which is readily measurable by conventional optical techniques[27–29], while the thermal conductance of supported graphene is modest due to its monolayer thickness and the suppression of in-plane phonon transport[30–32], compared to suspended graphene[33,34]. The addition of graphene dramatically increases the optical absorption to values measurable using conventional techniques, while not changing the thermal properties very much – thus serving as an ideal reference sample for the PCI measurement.

Using our self-referencing PCI technique, we measured the absorption coefficients of stoichiometric $Si_3N_4$ and silicon-rich $SiN_x$ ($x \sim 1$), determining that $Si_3N_4$ may be suitable as a candidate material for laser sails, with laser intensities approaching ~10 GW/m$^2$. Our self-referencing PCI technique opens up the possibility of measuring absorptivity in a variety of low-loss membranes without having to account for substrate effects.

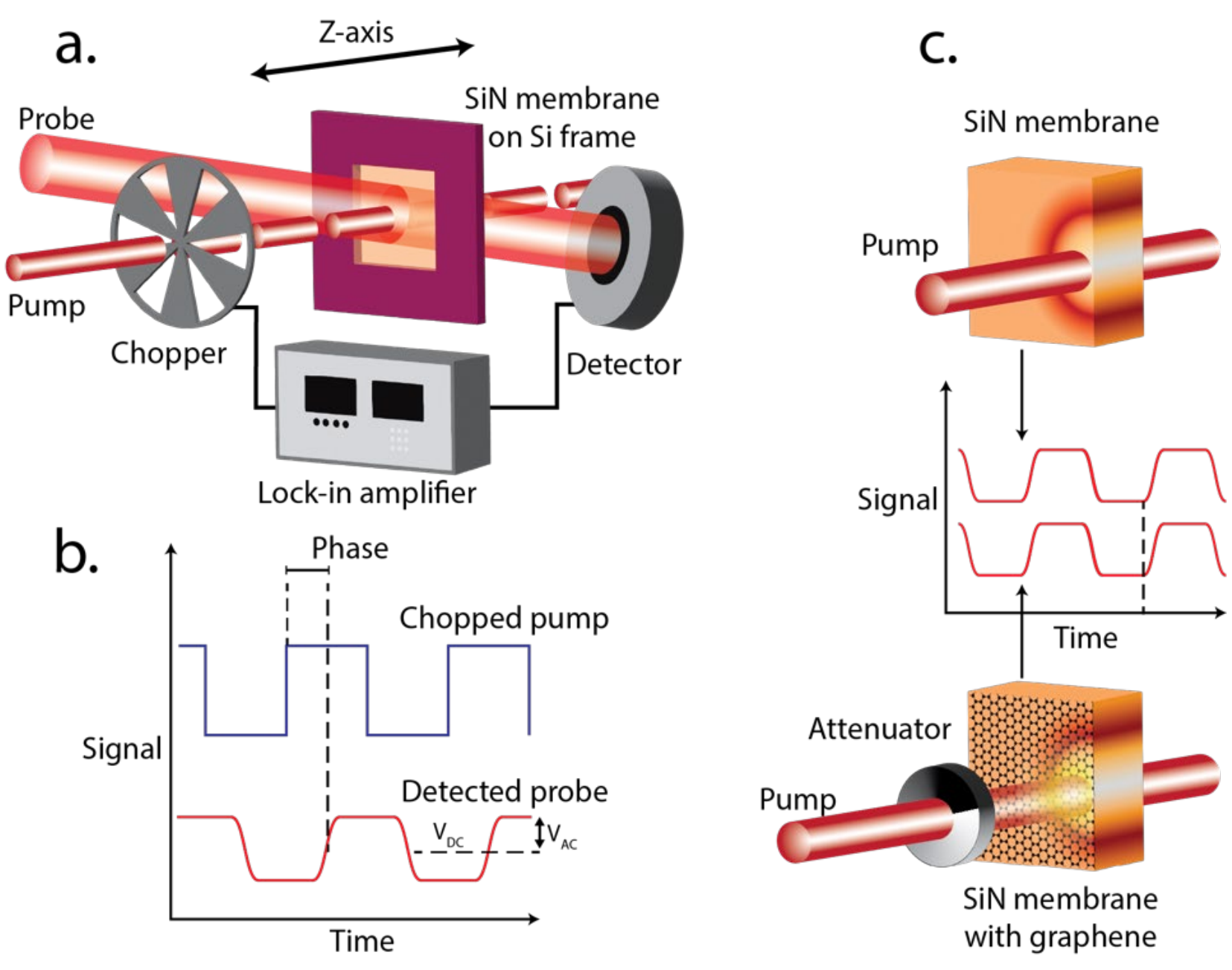

**Fig. 1**. (a) Schematic of the photothermal common-path interferometry (PCI) setup, along with (b) a visualization of the PCI signal (with AC and DC components) and phase resulting from the time delay between chopped light and detected probe intensity. (c) Transfer of a graphene monolayer onto a sample being characterized to increase optical absorption. The measurement of absorption then involves attenuating the pump until the same value of PCI signal is measured with graphene as the unattenuated measurement without graphene. In our experiments with silicon-nitride membranes, the addition of graphene did not significantly alter the PCI phase, indicating that the thermal conductance of the sample was not significantly altered.

## Self-referencing photothermal common-path interferometry

*Photothermal common-path interferometry (PCI)*

PCI measures the perturbation of a probe beam passing through a region of thermal lensing created by the absorption of a high-powered, chopped, continuous-wave pump laser (**Fig. 1a**)[4,22,35]. The chopped pump causes localized periodic heating in the vicinity of where the pump beam passes through the sample. The small increase in temperature in that region results in a change of refractive index via the thermo-optic effect, creating a thermal-lensing effect, and this change is measured using a probe laser at a different wavelength that is at an angle to the pump (**Fig. 1a**)[2,3,19,22–24]. Since the probe laser is bigger in diameter than the pump, the perturbed and unperturbed parts of the probe interfere with each other, leading to a diffraction pattern in the detector plane[4,19,22]. A diaphragm with a pinhole in front of the detector lets only the central peak of the diffraction pattern pass through, such that only the change in intensity of the central peak of the probe diffraction pattern is detected, and this carries information about the absorption of the material. Because the pump is chopped, the signal at the detector consists of AC and DC components ($V_{AC}$ and $V_{DC}$ respectively), and there is a delay, or phase difference, between the chopper and measured AC signal at the detector, which is related to the time constant of thermal dissipation (**Fig. 1b**)[2,22]. Typically in a PCI experiment, the peak $V_{AC}$ for a sample is used in calculations, which is achieved when the waists of the pump and the probe are both at the sample surface. To obtain this laser–sample configuration, the sample is moved in the *z*-direction (along the pump beam) until the characteristic peak in $V_{AC}$ is observed.

A PCI measurement does not directly provide the absolute absorptivity value; instead, the absorptivity $A$ (a unitless number between 0 and 1) must be obtained by translating the observed PCI signal. To first order, the absorptivity of the sample, $A$, can be related to $V_{AC}$, $V_{DC}$ and $P_{Pump}$ as:[4,35]

$$A = \frac{K \cdot V_{AC}}{P_{Pump} V_{DC}}, \quad (1)$$

where $P_{pump}$ is the power of the pump beam, and $K$ is a constant of proportionality sometimes referred to as a calibration or correction factor. We note that the definition of $K$ and the form of **Eqn. (1)** can vary across the literature [4,19–21,35]. In the present paper, $K$ has units of Watts, but there are certain works where both sides of **Eqn. (1)** have been normalized by sample thickness[4,21]. $K$ can depend on many variables, including the crossing angle, wavelength, and shape and size of the laser beams[21–24], and the sample's geometry and thermal properties, which include the heat capacity, thermal conductivity, thermo-optic coefficient, and coefficient of thermal expansion. Note, however, that thermal expansion affects the PCI signal via deformation in addition to the thermo-optic effect[22]. Therefore, $K$ must be determined for every new laser-beam setup, material, and geometry.

There exist various ways to determine $K$ (or a constant proportional to $K$) in the literature, but we found them challenging to apply to membranes. In one way to find $K$, a thin film of the sample in question can be grown on or transferred to a fused-silica substrate[2,21,25,26], or another substrate for which $K$ is known[21]; in our case, this would entail optimizing $Si_3N_4$ growth on or transfer onto a fused-silica substrate. $Si_3N_4$ growth is a non-trivial process requiring optimization of parameters such as gas flow, pressure, and temperature[25,36]. In addition, the form factor of a film on a substrate cuts off access to the back side of the film/membrane, which may be needed for future experiments such as the impact of dust on light sails[11]. We note that the approach involving film growth on a known substrate only works for films with thickness < 1-10 μm, because thermal lensing in thicker films (as opposed to the substrate underneath) can modify the PCI signal[2,21,35]. Another approach to determine $K$ is to perform a PCI measurement at a substantially different (often shorter) wavelength, where the optical absorptivity is larger and can be measured independently. However, this requires keeping the pump shape and size the same across different wavelengths[20,21]. $K$ has also been calculated theoretically[21,23,24], but the required multiphysics simulations have many input parameters resulting in many potential sources of error. A table of various methods to determine $K$ in the literature is available in **Supplemental Information S1**.

*Self-referencing technique for PCI*

We devised a new self-referencing PCI technique to calculate $K$, with the philosophy that the PCI reference sample should be as similar as possible to the sample being tested[22]. In this modality, we perform two PCI measurements: the first with the sample being investigated, and the second with the same (or identical) sample with a graphene monolayer transferred onto it. The use of monolayer graphene enhances the optical absorptivity of this reference sample to levels measurable by methods simpler than PCI (e.g., ellipsometry), while leaving its thermal conductance mostly unchanged, enabling the calculation of $K$ for a suspended membrane sample using **Eqn. (1)**. The thermal conductance is further discussed later in this manuscript.

To prepare the reference sample, we transferred CVD-grown monolayer graphene onto $Si_3N_4$ and $SiN_x$ membranes purchased from Norcada Inc.[37] (see **Supplemental Information S2** for membrane geometry), and then measured the absorptivity of the combined sample ($A_{ref}$) using variable-angle spectroscopic ellipsometry (see **Experimental Methods** and **Supplemental Information S3 and S4**) to be 1.5% ± 0.11% for the ~194-nm thick $Si_3N_4$ and 2.6% ± 0.16% for the ~2-µm thick $SiN_x$ at 1064 nm. These numbers are slightly different from the well-known ~2.3% absorptivity for suspended graphene[27–29] due to Fabry-Perot effects in the membranes. Precise thicknesses of the membranes were also calculated from ellipsometric measurements.

Then, we performed PCI measurements on the silicon-nitride membranes with and without graphene. For each PCI measurement, we translated the sample position along the z-axis (**Fig. 2a**) and recorded the PCI signal. The position where the PCI signal reaches its maximum (peak positions in **Fig. 2b** for $SiN_x$ and **Fig. 2c** for $Si_3N_4$ (dataset 1)) corresponds to the sample position where the pump and probe beams cross at the membrane; the values of $V_{AC}$ and phase at this position of maximal signal are then used for further analysis. We used a sufficiently powerful continuous-wave pump laser to obtain measurable PCI signals ($V_{AC}$, $V_{DC}$, and the phase) for the samples without graphene ($P_{pump,sample}$= ~2 W for $Si_3N_4$, and ~250 mW for $SiN_x$), and then attenuated the pump laser by many orders of magnitude ($P_{pump,ref}$= ~45 µW for $Si_3N_4$, and ~20 mW for $SiN_x$) to achieve similar $V_{AC}$ values between the sample and its graphene-coated reference.

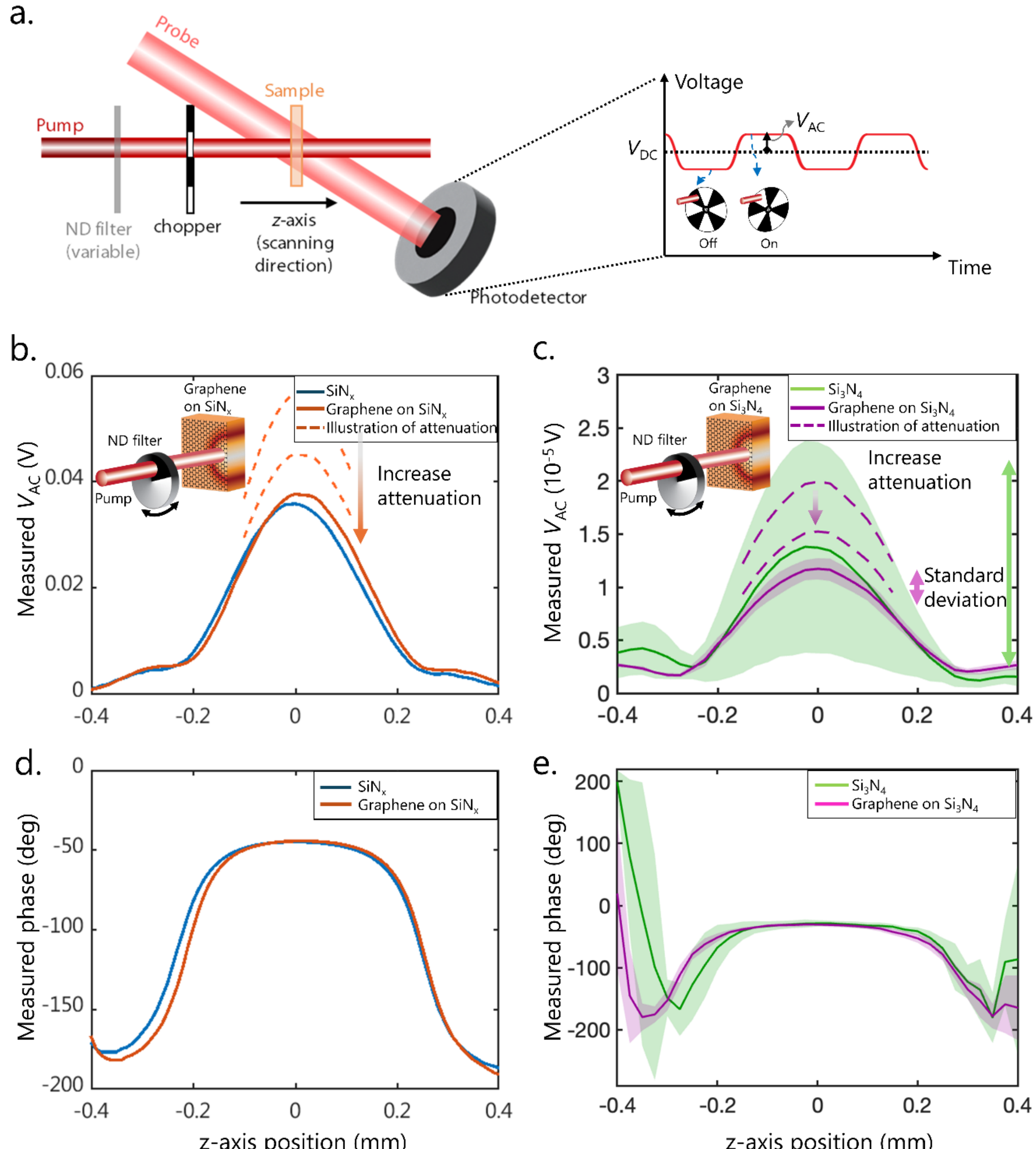

**Fig. 2**. (a) Side-view schematic of the PCI setup showing translation of the sample along the *z*-axis, and the AC and DC components of the detected signal. The sample is translated in the *z*-direction to find the peak of the AC signal which occurs when the pump waist is at the sample surface; (b) AC component of the detected probe intensity ($V_{AC}$) for the $SiN_x$ membrane with and without graphene. The pump intensity was manually attenuated using a variable ND-filter for the sample with graphene to obtain a similar $V_{AC}$ to that of $SiN_x$ alone (inset). Solid lines are the measured $V_{AC}$, while dashed lines represent the process of increasing attenuation to achieve similar $V_{AC}$ with and without graphene. Because of this manual attenuation process over many orders of magnitude of pump power, it was possible to bring the $V_{AC}$ of graphene-on-$SiN_x$ very close to that of $SiN_x$, but a slight difference in the $V_{AC}$ remained. That difference can be addressed via Eqn. (1) when calculating the absorption. (c) $V_{AC}$ for the $Si_3N_4$ membrane and $Si_3N_4$ with graphene (dataset 1), similarly obtained using a variable ND-filter. Due to the low loss of $Si_3N_4$, measurements spanned two datasets, and the average (solid lines) and standard deviation (shaded areas) of 5 measurements taken for $Si_3N_4$ and $Si_3N_4$ with graphene (dataset 1) are shown. Dashed lines illustrate the process of attenuation and do not represent actual measured data; (d, e) Phase between the chopped pump and detected probe intensities vs. the sample position for (d) the $SiN_x$ membrane and (e) $Si_3N_4$ membrane with and without graphene (dataset 1).

Because the absorptivity of the graphene-coated samples ($A_{ref}$) was already measured using ellipsometry, we used our PCI measurements and **Eqn. (1)** to calculate $K$, and then used $K$ to obtain absorptivities for the $Si_3N_4$ and $SiN_x$ samples. To obtain accurate numbers of average absorptivities and standard deviations for both stoichiometries, we took multiple measurements. Because of its low loss, $Si_3N_4$ measurements spanned two experimental datasets ($Si_3N_4$ (dataset 1) measurements used in **Figs. 2c** and **2e** and $Si_3N_4$ (dataset 2) measurements used in **Fig. 3a** and **supplemental information**), while a single dataset sufficed for $SiN_x$ (**Figs. 2b**, **2d**, and **3b**). We note that $Si_3N_4$ dataset 1 comprises data with high SNR but at fewer points on the membrane, while $Si_3N_4$ dataset 2 comprises data with lower SNR spanning many more points as shown in **Fig. 3a.**

We took 5 absorptivity measurements on $Si_3N_4$ (dataset 1), with results ranging from $2.47 \times 10^{-7}$ to $17 \times 10^{-7}$, corresponding to absorption coefficients ranging from $1.53 \times 10^{-2}$ cm$^{-1}$ to $10.56 \times 10^{-2}$ cm$^{-1}$. On $Si_3N_4$ (dataset 2), we conducted a surface absorptivity scan over 2601 points spanning an area of 0.5 mm × 0.5 mm, shown in **Fig. 3a**. We observed spikes in absorptivity that we believe to be dust particles[2,21] (similar features have been observed in other PCI measurements due to sample defects[21]), which were disregarded for the calculation of average absorptivity. The size of the spikes is ~80 μm, close to the pump diameter of 70 μm, implying that the dust particles are potentially much smaller than 80 μm. After discarding the outliers, we were left with ~1600 different spatial points, and we found the average absorptivity to be $(3.4 \pm 1.2) \times 10^{-7}$ in the ~194-nm membrane (i.e., $Si_3N_4$ (dataset 2)). Using the transfer-matrix method (see **Supplemental Information S5**), we converted this value to the absorption coefficient, which we found to be $(2.09 \pm 0.76) \times 10^{-2}$ $cm^{-1}$ at 1064 nm (the wavelength of our PCI pump). The calculation of the error bar is explained in **Supplemental Information S5**. Because the number of measurements and quality of data was different between $Si_3N_4$ datasets 1 and 2, we are unable to report an estimate for the absorption coefficient with an error bar. Nevertheless, based on the two sets of data, we estimate that the absorption coefficient of $Si_3N_4$ is between $1.5 \times 10^{-2}$ $cm^{-1}$ and $2.9 \times 10^{-2}$ $cm^{-1}$.

For comparison, we note that Land et al.[38] conducted nanomechanical absorption spectroscopy and reported an $Si_3N_4$ extinction coefficient ($\kappa$) of $1.8\times10^{-7}$ at 1064 nm, which corresponds to an absorption coefficient of $2.13 \times 10^{-2}$ $cm^{-1}$. At 1550 nm, Ji et al. and Luke et al. reported absorption coefficients of $3 \times 10^{-4}$ $cm^{-1}$ and $6.8 \times 10^{-3}$ $cm^{-1}$, respectively, using cavity ringdown spectroscopy, which does not directly observe loss in membranes and involves separating other loss mechanisms such as scattering in waveguides. We expect lower absorptivity at 1550 nm as compared to that at 1064 nm because there are no resonances until the mid-IR, though we also note (to our surprise) that Land et al. reported a higher absorption coefficient of $6 \times 10^{-2}$ $cm^{-1}$ at 1550 nm.

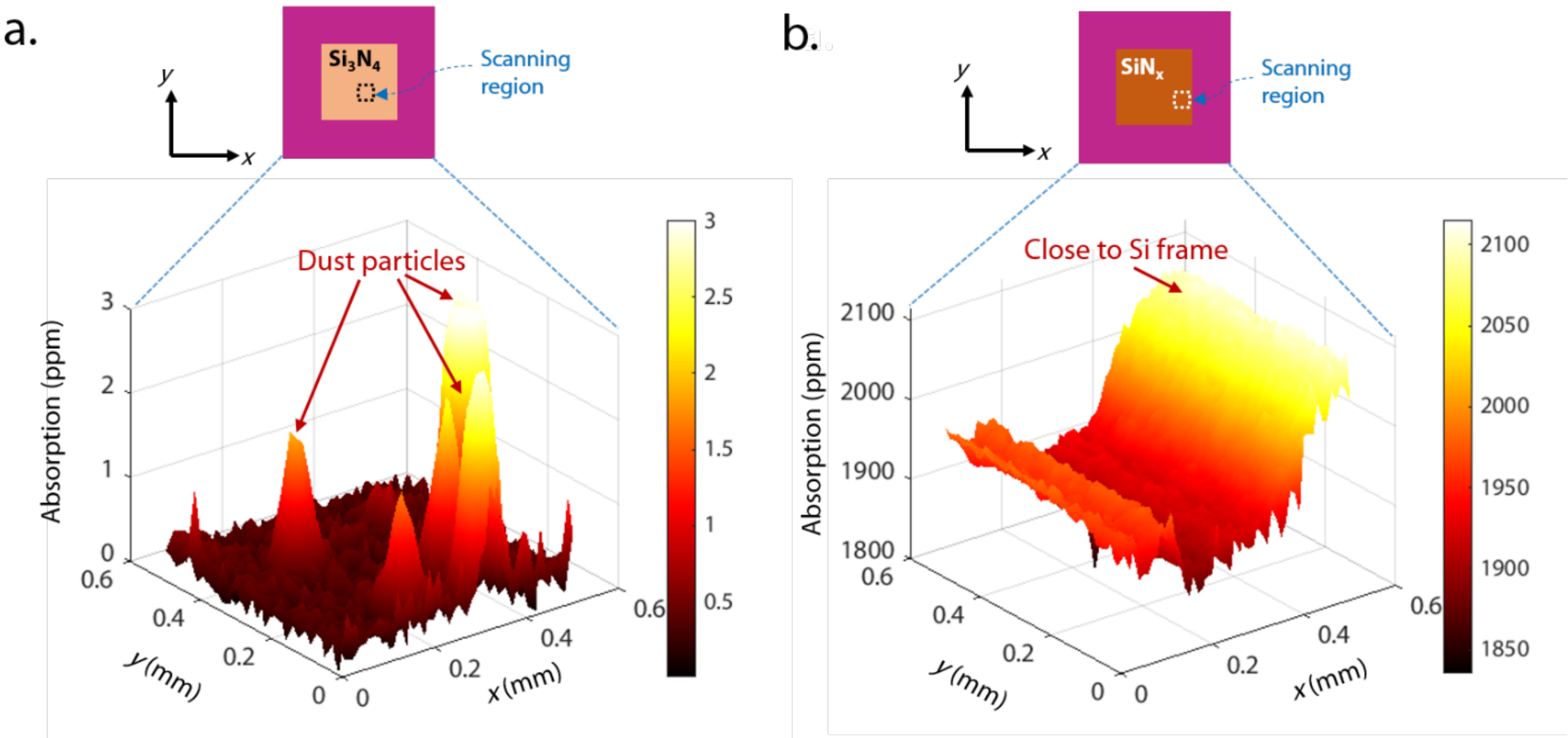

**Fig. 3**. 2D scans of absorptivity for (a) $Si_3N_4$ (dataset 2), and (b) $SiN_x$ ($x \sim 1$) membranes, in parts per million (ppm). The scanned area is a 0.5 mm × 0.5 mm square (dashed box on the membrane schematic) and visually presents spatial variability in absorption in the membranes. For the $Si_3N_4$ membrane, the sharp absorptivity peaks correspond to dust on the membrane, presenting a possible challenge for future laser sails. For the $SiN_x$ membrane, the measured absorption increases as the pump beam spot approaches the boundary of the membrane, such that a portion of the pump is absorbed in the Si frame. More information on the membrane dimensions is available in **Supplemental Information S2**.

For the $SiN_x$ membrane, we conducted measurements over 2601 points spanning an area of 0.5 mm x 0.5 mm, as we did with $Si_3N_4$ (dataset 2). We observed an increase in absorptivity when the pump laser beam was close to the Si frame (**Fig. 3b**) and excluded these points of high absorption from the average absorptivity calculation. We calculated the average absorptivity to be $(1.94 \pm 0.12) \times 10^{-3}$ for the ~2-μm $SiN_x$ membrane, corresponding to an absorption coefficient of $7.94 \pm 0.50$ $cm^{-1}$ (**Supplemental Information S5**). This is close to the reported value of $(6.9 \pm 0.7)$ $cm^{-1}$ using PCI and cavity round-trip measurements by Steinlechner et al.[3], and is on the same order of magnitude of loss reported for various stoichiometries of PECVD-grown $SiN_xH_y$ measured using PCI[39]. We note that in Steinlechner et al.[3], this number is reported for "Low-stress 2 μm $Si_3N_4$ membranes", which we understand to actually be $SiN_x$ membranes similar to the ones we study here (x∼1).

**Fig. 3** demonstrates potential challenges for laser sails relating to the variability of absorption across the membrane, both due to intrinsic differences in the material and due to increases in absorptivity caused by dust on the membrane (either due to contamination during fabrication or from space itself). The effect of space dust, and absorption variability more broadly, has been previously considered in the context of laser sails[11]. In terms of integrity of a hypothetical laser sail, the low absorptivity of stoichiometric $Si_3N_4$ is encouraging, though additional measurements are needed to characterize the temperature dependence of the absorption coefficient relevant in thermal runaway processes[10], and care must be taken to avoid variations in silicon nitride stoichiometry that lead to higher absorptivity. Simplified calculations of the temperature of a simple $Si_3N_4$ membrane sail under laser illumination and assuming temperature-independent absorptivity can be found in **Supplemental Information S7**. We calculated the equilibrium temperature of a semi-infinite $Si_3N_4$ slab under 10 GW $m^{-2}$ of illumination. For the absorption coefficient ranging from $1.5 \times 10^{-2}$ $cm^{-1}$ to $2.9 \times 10^{-2}$ $cm^{-1}$, we calculated the equilibrium temperature to be from ~698 K to ~939 K, lower than the decomposition temperature of $Si_3N_4$ at 1500 K – 1900 K[40,41].

## Validity of the self-referencing technique

A key assumption in our self-referencing PCI technique is that the pump-beam-induced thermal-lensing effect within the sample is similar to that within the reference. This assumption can be validated using the PCI phase, because the phase depends on the material's thermal properties and is thus a good method of comparing thermal lensing between samples[2,21]. In our self-referenced PCI experiments, the addition of graphene to a sample did not significantly change the measured PCI phase (**Figs. 2d, 2e**), indicating that the heat generated during optical absorption at the graphene is quickly transferred to the sample underneath and the overall thermal conductance is dominated by the sample itself, with only a minor contribution from the graphene.

This observation is supported by individually considering the thermal conductances (which depend on sample thicknesses and the thermal conductivities) of graphene and silicon nitride membranes. We measured the in-plane thermal conductivity of $Si_3N_4$ and $SiN_x$ to be approximately 18 $W \cdot m^{-1} \cdot K^{-1}$ and 10.3 $W \cdot m^{-1} \cdot K^{-1}$, respectively, using frequency domain thermoreflectance (see *Experimental Methods*). We did not measure the thermal conductivity of our graphene directly, but we do not expect it to be more than roughly 1000 $W \cdot m^{-1} \cdot K^{-1}$, a conservative estimate based on reported thermal conductivities of supported graphene in literature[30–32]. We note that the thermal conductivity of supported graphene is much lower than that of freestanding graphene (up to 5000 $W \cdot m^{-1} \cdot K^{-1}$) for two reasons: (a) suppression of flexural modes that are present in freestanding graphene[31,32], and (b) leakage of phonons across the graphene-membrane interface [30,31]. Thermal conductance is then directly proportional to the product of thermal conductivity and sample thickness; in this case the monolayer thickness of graphene (0.335 nm) leads to an order of magnitude lower thermal conductance of graphene (proportional to $0.3 \text{ nm} \times 1000 \text{ W m}^{-1} \text{ K}^{-1} = 3 \times 10^{-7} \text{ W K}^{-1}$) than that of the $Si_3N_4$ membrane (proportional to $200 \text{ nm} \times 18 \text{ W m}^{-1} \text{ K}^{-1} = 3.6 \times 10^{-6} \text{ W K}^{-1}$).

## Conclusion

Characterization of optical absorption of low-loss materials is important for applications in on-chip photonics, optical components in sensitive experiments, and (most-relevant to this paper) laser-light sails. Here, we demonstrated a self-referencing approach to photothermal common-path interferometry (PCI), wherein the transfer of monolayer graphene onto a given low-loss sample significantly increases its absorptivity to create a reference for the PCI technique. For all membranes we studied, the addition of graphene did not significantly affect the thermal properties of the sample underneath, preserving the validity of PCI. Based on two sets of measurements, we estimated the absorption coefficient of stoichiometric $Si_3N_4$ to be roughly between $1.5 \times 10^{-2}$ cm$^{-1}$ and $2.9 \times 10^{-2}$ cm$^{-1}$ at 1064 nm. For a non-stoichiometric silicon nitride ($SiN_x$) sample, we measured the absorption coefficient to be approximately 8 cm$^{-1}$. The absorption coefficient of stoichiometric $Si_3N_4$ is sufficiently small to enable light sails at incident intensities approaching ~10 GW/m$^2$ assuming no runaway thermal processes, although care must be taken to avoid variations in its stoichiometry. Our self-referencing PCI technique using monolayer graphene can be applied to most suspended membranes or more-complex structures, and is a promising way to evaluate low-loss materials.

## Acknowledgements

We thank Alexei Alexandrovski, Jessica Steinlechner, and Ross Johnston for helpful discussions and reference information. This work was supported by the National Science Foundation (1750341), Office of Naval Research (N00014-20-1-2297), the UW-Madison Research Forward Initiative via the Wisconsin Center for Semiconductor Thermal Photonics, and the UW-Madison Student Research Grants Competition. We gratefully acknowledge the use of facilities at the UW-Madison Soft Materials Characterization Lab, part of the UW-Madison Wisconsin Centers for Nanoscale Technology (wcnt.wisc.edu), which is partially supported by the NSF through the University of Wisconsin Materials Research Science and Engineering Center (DMR-2309000).

## Experimental methods

<u>*Sample details*</u>

We measured the absorptivity in ~194-nm thick $Si_3N_4$ and ~2-µm thick $SiN_x$ membranes (purchased from Norcada Inc., Edmonton, AB, Canada), suspended on 200 µm thick silicon frames. The thicknesses of the membranes were calculated from ellipsometric measurements. In the in-plane direction, membranes of both stoichiometries had the same dimensions with a silicon frame of 10 mm x 10 mm and a freestanding membrane area of 5 mm x 5 mm (see **Supplemental Information S2**).

*PCI experimental setup*

The PCI setup comprised a 1064 nm pump laser (YLR-10-LP, IPG Photonics) and a 633 nm probe laser (JDSU 1122P HeNe laser). The pump beam was chopped at ∼390 Hz and its power measured by a thermopile detector (Thorlabs S310C), and the modulated probe signal measured with a detector (DET10A Si detector, Thorlabs, Inc.), connected to a lock-in amplifier (SRS SR810) (**Fig. 1a**).

The pump and probe beams were set to cross each other at their beam waists. The sample was moved in the *z*-direction until its surface was in the sample plane as the beam waists, indicated by a characteristic peak in the AC signal[22] (**Fig. 2**). All subsequent measurements for a given sample were conducted at the *z*-position thus obtained.

To obtain similar AC voltage values for the graphene-coated reference and the sample to be measured, the pump laser power was appropriately attenuated for the former. In normal PCI operation the pump is attenuated with a half-wave plate and polarizer; for powers lower than 1 mW, we used an additional neutral density (ND) filter with optical density (OD) of 0.9. In this power regime, a power meter (Thorlabs S130VC) was placed between the chopper and the sample to note the power. 1-D *z*-scan (longitudinal direction) PCI signals (**Fig. 2**) as well as 0.5 mm×0.5 mm 2D maps were acquired for each sample (**Fig. 3**).

For each stoichiometry, upon adjusting the pump for similar AC signal values with and without graphene, we noted the ratio of the respective pump powers required, to calculate absorptivity values. To account for surface variations, we measured the PCI signal at multiple spatial points on each membrane. We conducted measurements at 5 points on $Si_3N_4$ (dataset 1) and at 2601 points for $Si_3N_4$ (dataset 2). For $SiN_x$, a single data set of 2601 points sufficed; a relatively good signal was measured because of the higher absorptivity of $SiN_x$. We discarded anomalous data such as possible specks of dust and increasing absorptivity close to the frame.

*Transfer of graphene onto SiN membranes, and ellipsometry characterization*

Graphene was transferred onto the $Si_3N_4$ and $SiN_x$ membranes using a wet transfer method. Polymethyl methacrylate (PMMA) was spin-coated onto CVD graphene grown on a copper foil (obtained from Grolltex Inc., San Diego, CA). The Cu foil was etched away in $FeCl_3$. The graphene was then transferred onto the membranes and baked at 60 °C to ensure good adhesion and the removal of water between graphene and the membrane. The PMMA was removed by an acetone bath at 60 °C.

The absorptivities of the graphene-coated $Si_3N_4$ and $SiN_x$ membranes were calculated using variable angle spectroscopic ellipsometry (J.A. Woollam V-VASE). First, the thickness and refractive index ($n$) of the $Si_3N_4$ and $SiN_x$ membranes were obtained from models fitted to the ellipsometric parameters psi $\Psi$ and delta $\Delta$. Then, these models were used as substrates for ellipsometric data of samples with graphene, and their absorptivities calculated using J.A. Woollam's WVASE software (which uses the transfer matrix method). **Supplemental sections S3 and S4** list out more details about the ellipsometry and corresponding fitting.

*Frequency-domain thermoreflectance measurements of $Si_3N_4$ and $SiN_x$ thermal conductivity*

We used frequency-domain thermoreflectance (FDTR) to measure the thermal conductivities of the $SiN_x$ and $Si_3N_4$ membranes. We deposited a ~100 nm Au film, which has a large temperature coefficient of thermoreflectance[42], on the membranes using electron beam evaporation. The pump (488 nm, Coherent Genesis MX 1W) and probe (532 nm Coherent OBIS LX 20 mW) beams were focused through a 20× infinity-corrected objective lens to achieve 6.3 and 5.9 μm spot sizes at the transducer surface. The pump and probe powers were fixed at 3 mW and 2.2 mW, respectively, in order to limit the temperature-rise at the sample surface to < 1 K[43]. Literature values for the volumetric heat capacities of $SiN_x$[44] and $Si_3N_4$[45] were used for extracting the thermal boundary conductance at the Au/membrane interfaces and the in-plane thermal conductivities of the membranes. Using FDTR, we fit both the in-plane and cross-plane thermal conductivities of thin-film samples. In our FDTR measurements, we were not sensitive to the cross-plane thermal conductivity of the membrane - the thermal penetration depth of the heating laser was larger than the membrane thicknesses for a significant part of the applied frequency range, causing the response of the probe beam to be governed by only the in-plane thermal transport. The cross-plane thermal conductivities were measured separately in regions where the membrane is supported by a silicon substrate, and confirmed that we were not sensitive to this value in regions where the membrane was suspended.

Supplemental information for

# Self-referencing photothermal common-path interferometry to measure absorption of $Si_3N_4$ membranes for laser-light sails

Tanuj Kumar[1‡], Demeng Feng[1‡], Shenwei Yin[1], Merlin Mah[2], Phyo Lin[2], Margaret A. Fortman[3], Gabriel R. Jaffe[3], Chenghao Wan[1,4], Hongyan Mei[1], Yuzhe Xiao[1,5], Ron Synowicki[6], Ronald J. Warzoha[7], Victor W. Brar[3], Joseph J. Talghader[2], Mikhail A. Kats[1♦]

[1]Department of Electrical and Computer Engineering, University of Wisconsin–Madison, Madison, WI 53706, USA

[2]Department of Electrical and Computer Engineering, University of Minnesota–Twin Cities, MN 55455, USA

[3]Department of Physics, University of Wisconsin–Madison, Madison, WI 53706, USA

[4]Department of Electrical Engineering, Stanford University, Stanford, CA 94305, USA

[5]Department of Physics, University of North Texas, Denton, TX, 76203, USA

[6]J. A. Woollam Co. Inc., 645 M St Suite 102, Lincoln, NE 68508, USA

[7]Department of Mechanical Engineering, United States Naval Academy, Annapolis, MD 21402, USA

‡ - equal contribution

♦ - corresponding author. Email: mkats@wisc.edu

## S1. Summary of methods to determine the "correction factor" $K$ in the literature

**Table S1.** Comparison of methods to determine $K$ in the literature

| **Method** | **Theoretical/ Experimental** | **Sample used** | **Notes** | **References (supplementary)** |
|---|---|---|---|---|
| Model heat distribution and probe distortion | Theoretical | Suprasil 311, KU-1 | Complicated, many variables to take care of | [S1,S2] |
| Model probe distortion then (semi) numerically solve | Theoretical | AlGaAs on fused silica, AlGaAs on sapphire | | [S3] |
| Grow thin-film on bulk fused silica | Experimental | AlGaAs film on fused silica | Not applicable to membranes | [S4–S7] |
| PCI measurement with known loss at different wavelength | Experimental | GaAs/AlGaAs coating on fused silica, Germanium | Beam shape and size must be same at shorter wavelength. Loss at different wavelength not always known | [S3,S8] |
| Increase loss by doping | Experimental | $LiNbO_3$ | | [S9] |
| Reference sample by addition of monolayer graphene | Experimental | $Si_3N_4$/$SiN_x$ (x~1) membranes | Reference sample has similar thermal properties as but orders-of-magnitude higher absorbance than sample being tested, which bypasses sample and setup variabilities | This work |

## S2. Membrane geometry

We obtained ~200 nm thick $Si_3N_4$ and ~2 $\mu$m thick silicon-rich $SiN_x$ ($x \sim 1$) membranes from Norcada Inc. (Edmonton, AB, Canada). Both membranes were obtained mounted on 200 $\mu$m thick, 10 mm × 10 mm Si frames, with a 5 mm × 5 mm area of suspended membrane in the middle (**Fig. S1**).

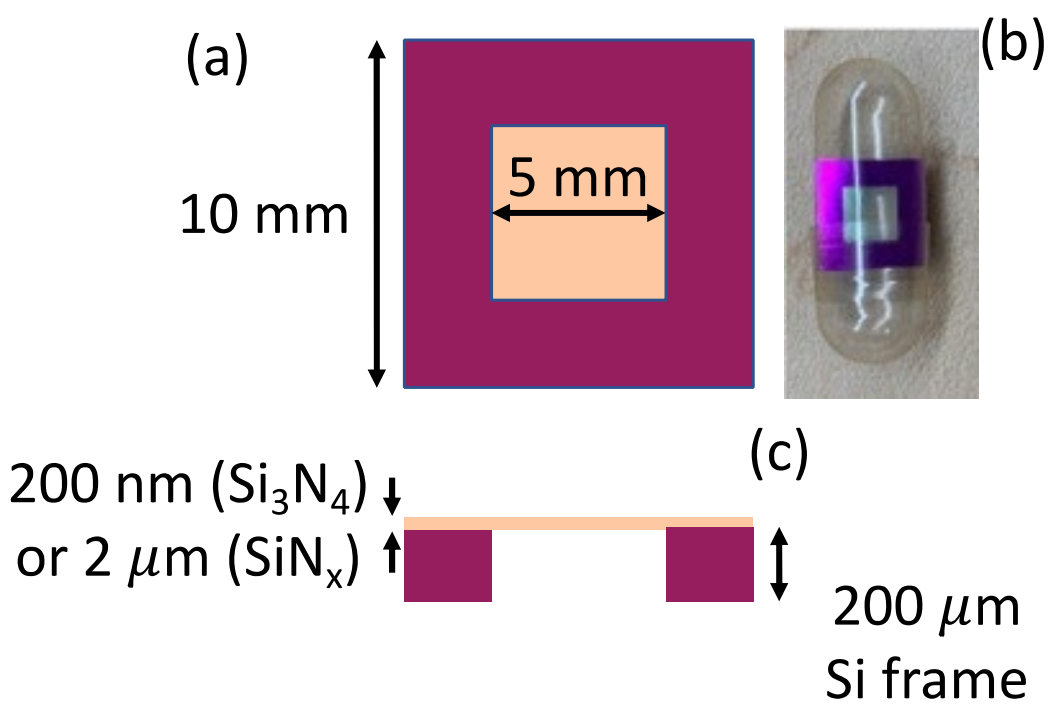

**Fig. S1**. (a) Front-view schematic of the membranes, (b) front-view picture of a $Si_3N_4$ membrane enclosed in a clear capsule, (c) side-view schematic of the membranes showing the suspended part of the membranes in the middle.

## S3. Characterization of thickness and refractive index of the $Si_3N_4$ membrane via variable-angle spectroscopic ellipsometry

We performed variable-angle spectroscopic ellipsometry measurements on $Si_3N_4$ and $SiN_x$ membranes mentioned in the main text, over a wavelength range of 300 – 1500 nm. We used a J. A. Woollam V-VASE ellipsometer for measurements, and used several different oscillator models to fit the experimental data. In this section, we discuss in detail the data and analysis for $Si_3N_4$, but similar analysis and conclusions apply to $SiN_x$ ($x \sim 1$).

We used three different models to fit the ellipsometry data: Cauchy model with Urbach tail, Tauc-Lorentz model, and Cody-Lorentz model, of which the last 2 satisfy the Kramers-Kronig relations. We used the model expressions for these oscillator models from J.A. Woollam, Inc.'s handbook on using WVASE ellipsometry fitting software[S10]. The expressions, together with the resulting fitting parameters, including the thicknesses, are shown in **Table S2**. All those models fit the experimental ellipsometry data (300 – 1500 nm) well (as shown in **Fig. S2**).

**Table S2.** Fitted expressions of the complex relative permittivity, and the resulting membrane thickness from different models

| Model | Complex relative permittivity $\epsilon_r$ [a] | Membrane thickness |
|---|---|---|
| Cauchy | $n = A + \frac{B}{\lambda^2} + \frac{C}{\lambda^4}$ [b]<br>$\kappa = \alpha \cdot \exp(\beta(E - \gamma))$ [c]<br>$\epsilon_r = (n + i \cdot \kappa)^2$ [d]<br>where $A = 1.9906, B = 0.014928, C = 0.0004, \alpha = 0.27783, \beta = 1.6526, \gamma = 6.199$. | 194.3 nm |

<table>
<tr><td>Tauc-Lorentz</td><td>$\epsilon_2(E) = \begin{cases} \frac{1}{E}\cdot\frac{AE_oC(E-E_g)^2}{(E^2-E_o^2)^2+C^2E^2} & (E \geq E_g) \\ 0 & (E < E_g) \end{cases}$<br>$\epsilon_1(E) = 1 + \frac{A_p}{E_p^2-E^2} + \frac{2}{\pi}\mathcal{P}\int_0^\infty \frac{\xi\epsilon_2(\xi)}{\xi^2-E^2}d\xi$ [e]<br>$\epsilon_r = \epsilon_1 + i\cdot\epsilon_2$<br>where $A = 0.69751, E_o = 5.7188, C = 0.64145, E_g = 0, A_p = 228.51, E_p = 8.9649$.</td><td>194.4 nm</td></tr>
<tr><td>Cody-Lorentz</td><td>$\epsilon_2(E) =$<br>$\begin{cases} \frac{E_g+E_t}{E}\cdot G(E_g+E_t)\cdot L(E_g+E_t)\cdot\exp\left(\frac{E-E_g-E_t}{E_u}\right) & (0 < E \leq E_g+E_t) \\ G(E)\cdot L(E) = \frac{(E-E_g)^2}{(E-E_g)^2+E_{pCL}^2}\cdot\frac{AE_o\Gamma E}{(E^2-E_o^2)^2+\Gamma^2E^2} & (E > E_g+E_t) \end{cases}$<br>$\epsilon_1(E) = 1 + \frac{A_p}{E_p^2-E^2} + \frac{2}{\pi}\mathcal{P}\int_0^\infty \frac{\xi\epsilon_2(\xi)}{\xi^2-E^2}d\xi$<br>$\epsilon_r = \epsilon_1 + i\cdot\epsilon_2$<br>where $A = 0.73147, E_o = 5.7237, \Gamma = 0.67386, E_g = 0.97139, E_{pCL} = 1, E_t = 0, E_u = 0.5, A_p = 229.12, E_p = 8.9752$.</td><td>194.4 nm</td></tr>
<tr><td colspan="3">Note:<br>[a]Variables in blue in the $\epsilon_r$ column are the fitted parameters in each model.<br>[b]$\lambda$: free-space wavelength. All $\lambda$ in this table have the unit of $\mu$m.<br>[c]$E$: photon energy. All $E$ in this table have the unit of eV.<br>[d]$i$: all $i$ in this table refer to imaginary unit ($i^2 = -1$).<br>[e]$\mathcal{P}$ denotes the Cauchy principal value.</td></tr>
</table>

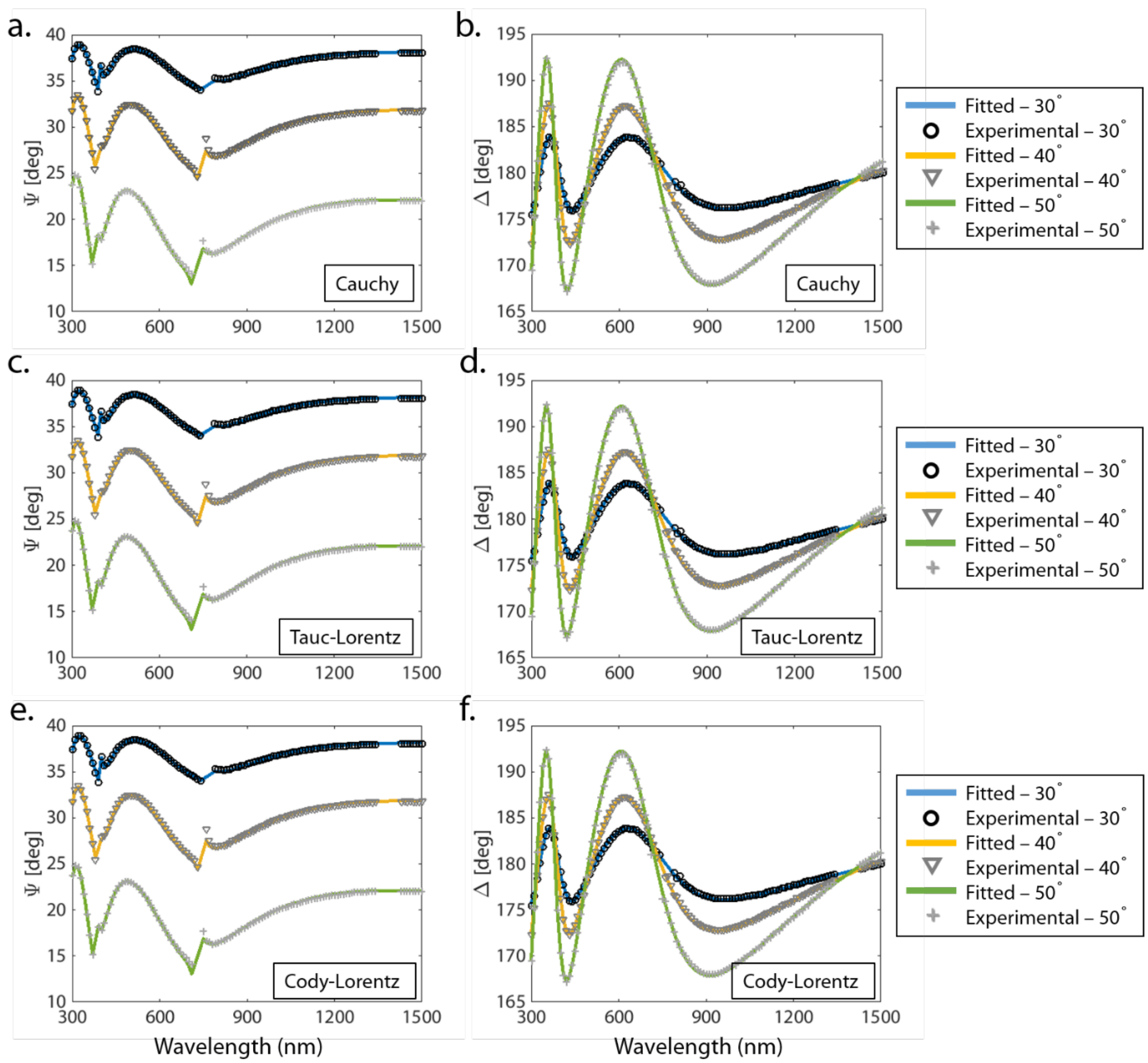

**Fig. S2**. Ψ and Δ data from the ellipsometry measurements at different incident angles, together with the model fitting results using the Cauchy model (a-b), the Tauc-Lorentz model (c-d), and the Cody-Lorentz model (e-f). All 3 models fit the experimental data well in this wavelength range.

We calculated the complex refractive indices from these 3 fitted models using the equation $n + i \cdot \kappa = \sqrt{\epsilon_r}$, where $n$ is the real part of the refractive index, and $\kappa$ is the extinction coefficient. Although those models agree on $n$ values across the whole wavelength range, these models give $\kappa$ values that differ by several orders of magnitude (**Fig. S3**). This is because our membrane is too thin and the extinction coefficient too small, leading to a very short optical path inside the membrane. Consequently, even a relatively big change in $\kappa$ does not lead to a significant change in the ellipsometric parameters, and these findings necessitated the use of PCI to characterize the extinction coefficient of $Si_3N_4$. Furthermore, we note that all 3 models give $n = 2.004$ at 1064 nm for $Si_3N_4$, and the thickness of the membrane is ~194 nm. Those values were used in the PCI data analysis to determine the absorption coefficient of $Si_3N_4$ (see **Supplemental Information S5**).

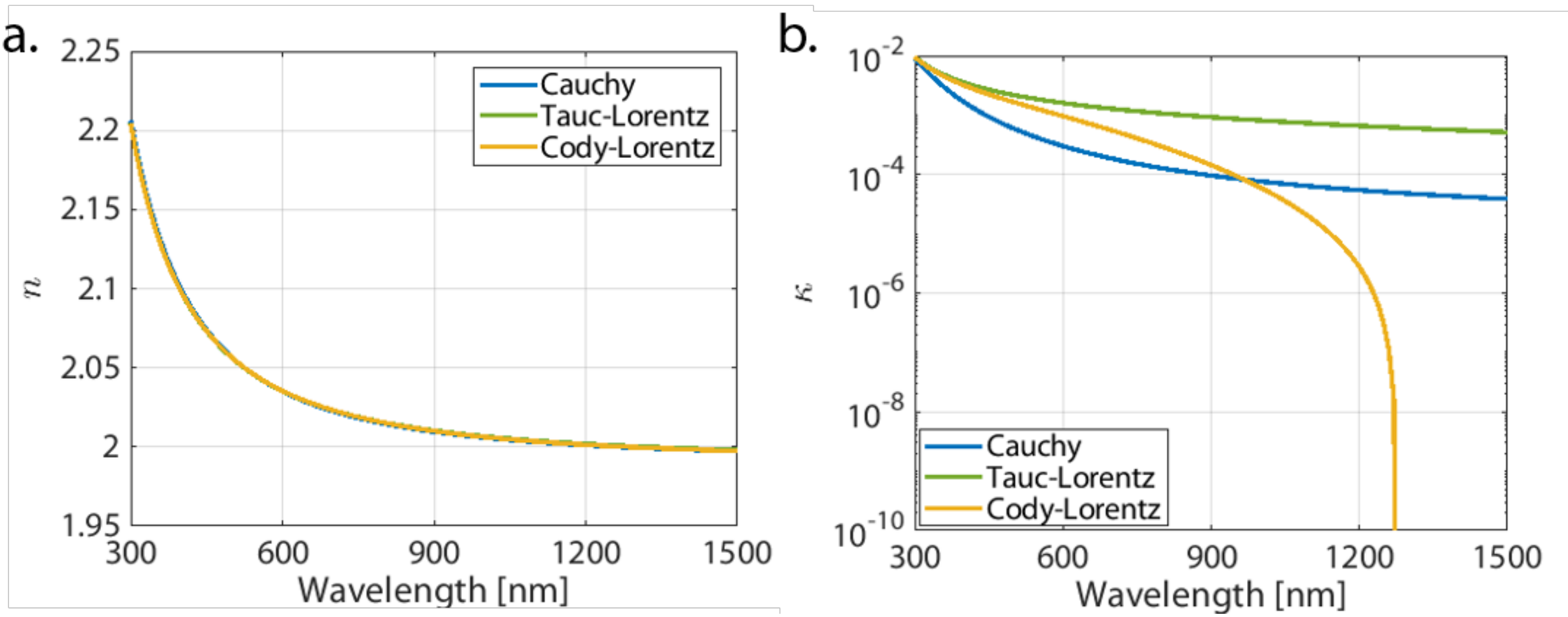

**Fig. S3**. Real (a) and imaginary (b) parts of the refractive indexes from fitted models. Although different models agree on $n$ values, these models give different $\kappa$ values, and therefore another characterization method with higher precision is needed to characterize $\kappa$ for $Si_3N_4$.

We observed similar behavior for $SiN_x$ ($x \sim 1$). Specifically, at 1064 nm, different models give $n = 2.147$, but give $\kappa$ values that differ several orders of magnitude. All models show the thickness of the $SiN_x$ membrane to be ~ 2.01 $\mu$m. Those values will be used in the PCI data analysis to determine the absorption coefficient of $SiN_x$ (see **Supplemental Information S5**).

## S4. Absorptivity measurements of reference samples using variable-angle spectroscopic ellipsometry

We determined the absorptivity of graphene-on-$Si_3N_4$ and graphene-on-$SiN_x$ membranes using variable angle ellipsometry. Between samples with and without graphene, we observed enough difference in the ellipsometric parameters psi ($\Psi$) and delta ($\Delta$) to conclude that graphene absorbance was observable via variable angle ellipsometry.

*$Si_3N_4$*

Our model to fit the ellipsometric data included a $Si_3N_4$ membrane coated by a 0.335 nm layer of graphite with a layer of water in between the two (**Fig. S4**). This is similar to the model used in Kravets et al.'s work on the ellipsometry of graphene[S11]; they fix the thickness of graphene to 0.335 nm and then fit the thickness of the water layer. **Figs. S4a** and **b** show the comparison of $\Psi$ and $\Delta$ for samples with and without graphene; while measurements were taken at 3 different spatial positions on each sample at 65°, 70°, and 75° angles of incidence, data from only one measurement of each sample at 75° angle-of-incidence is shown in **Fig. S4** for clarity. A simple Drude oscillator model for the graphene also provided good fits, but we used a graphite model available in J.A. Woollam's WVASE library (**Table S3**) for its accuracy in capturing increasing absorption of graphene towards UV wavelengths[S11]. This fit resulted in a mean square error of fitting of ~1.1-2.3, and an absorptivity of 1.5% $\pm$ 0.11% for the graphene-on-$Si_3N_4$ sample at 1064 nm. We verified the accuracy of this model by noting that the absorptivity of the 0.335 nm layer of graphite alone was calculated to be 2.75% at 1064 nm, slightly higher than the well-known graphene absorbance of 2.3%.

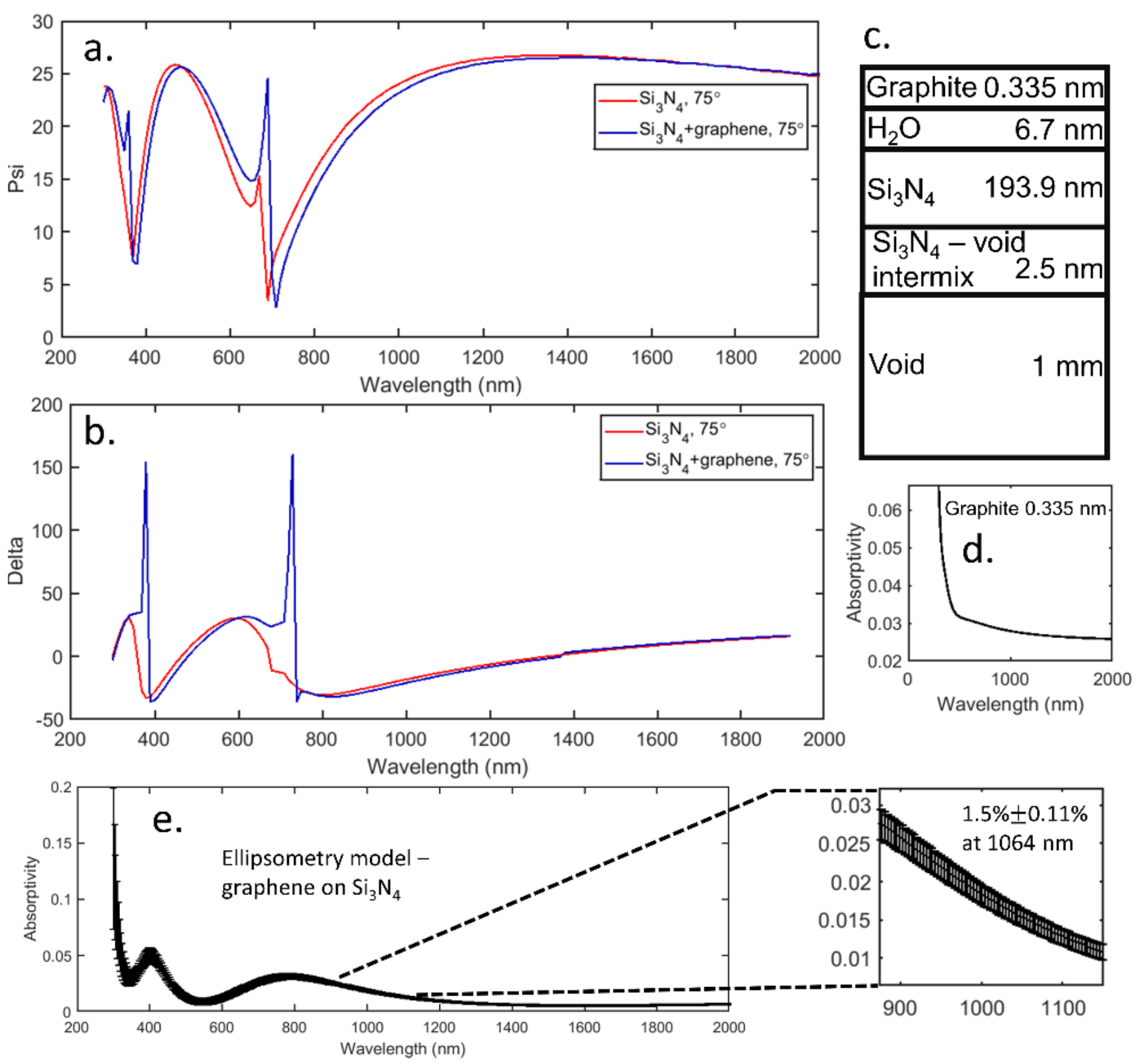

**Fig. S4**: Comparison of the measured (a) psi and (b) delta of $Si_3N_4$ membranes with and without graphene, (c) model used to fit ellipsometric data of samples with graphene, (d) calculated absorptivity of a 0.335 nm layer of graphite using a model from J.A. Woollam's library[S10], (e) calculated absorptivity of the model for the sample with graphene (1.5%±0.11%).

**Table S3.** Oscillators and parameters for the model describing graphite, used as a component of our model for graphene deposited on the $Si_3N_4$ membrane

| Oscillator | Parameter and value | | |
|---|---|---|---|
| Drude<br>$\epsilon = -\frac{A_n Br_n}{E^2 + iBr_n E}$ | $A_n$ = 15.048 | Br = 4.8914 | |
| Gaussian<br>$\epsilon = \epsilon_{n1} + i\epsilon_{n2}$<br>Where<br>$\epsilon_{n2} = A_n e^{-\left(\frac{E-E_n}{\sigma}\right)^2} - A_n e^{-\left(\frac{E+E_n}{\sigma}\right)^2}$,<br>$\epsilon_{n1} = \frac{2}{\pi}\mathcal{P}\int_0^\infty \frac{\xi\epsilon_{n2}(\xi)}{\xi^2-E^2}d\xi$,<br>and $\sigma = \frac{Br_n}{2\sqrt{\ln(2)}}$ | $A_n$ = 2.8924 | $E_n$ = 2.0716 | Br = 2.1145 |
| Gaussian | $A_n$ = 6.9623 | $E_n$ = 4.5182 | Br = 0.8383 |
| Gaussian | $A_n$ = 2.5276 | $E_n$ = 5.2156 | Br = 1.8065 |
| Gaussian | $A_n$ = 3.2927 | $E_n$ = 3.7064 | Br = 1.2689 |

*SiN$_x$*

Our process to model graphene on SiN$_x$ was similar to that with Si$_3$N$_4$ as detailed above, with the exception that SiN$_x$ and graphene-on-SiN$_x$ ellipsometry was done only at angles of 70°, 75° because the data at 65° was noisy. **Fig. S5** shows ellipsometric data comparison, model used and absorbance of the SiN$_x$ membrane with graphene on it (2.6% ± 0.16% at 1064 nm).

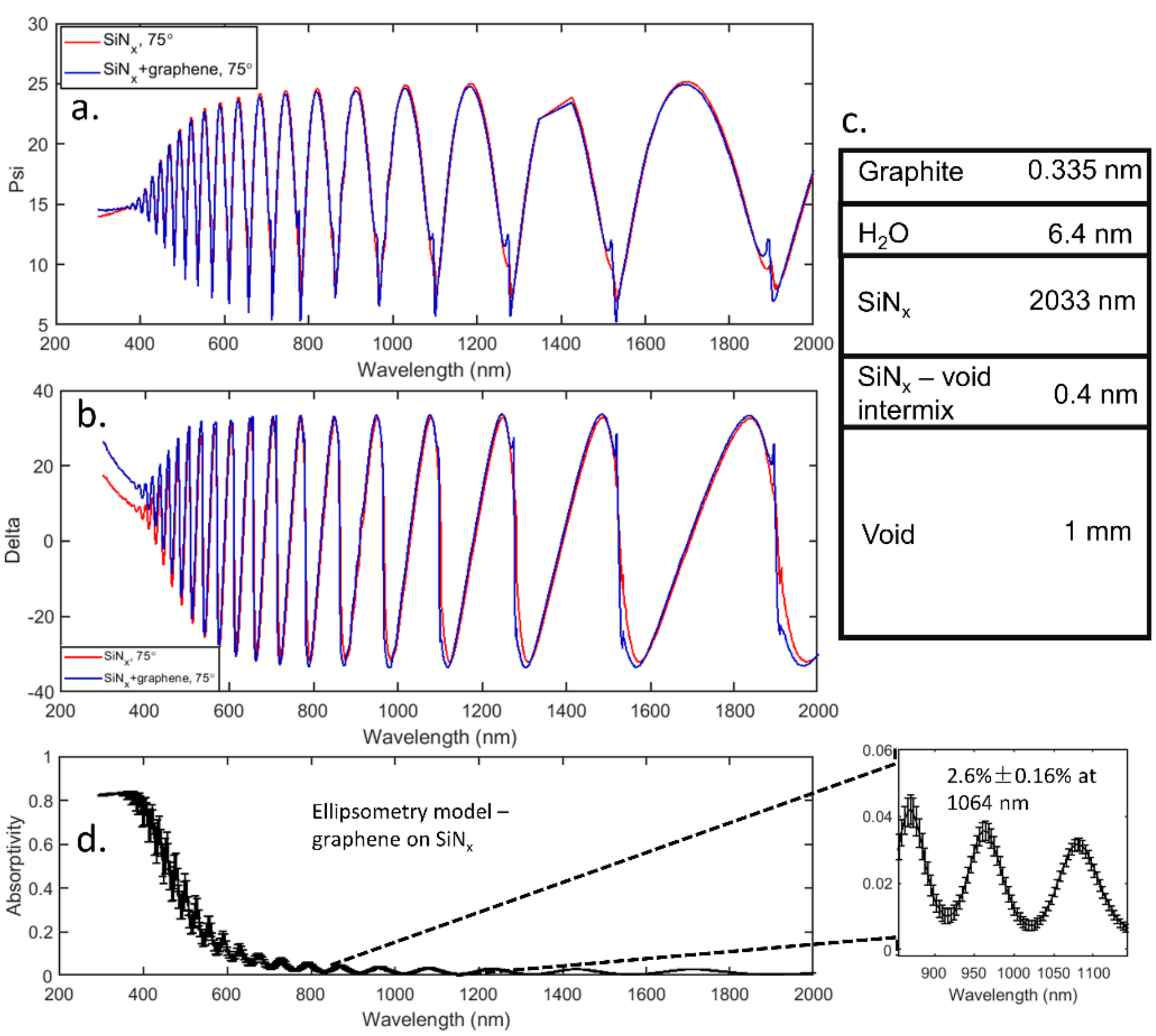

**Fig. S5**: Comparison of the measured (a) psi and (b) delta of SiN$_x$ membranes with and without graphene, (c) model used to fit ellipsometric data of samples with graphene, (d) calculated absorptivity of the model for sample with graphene (2.6% ± 0.16% at 1064 nm).

## S5. PCI data analysis

*Membrane absorptivity calculations for Si$_3$N$_4$ (dataset 2) and SiNx*

We used the self-referencing PCI method discussed in the main text to calculate the absorptivity of the Si$_3$N$_4$ and SiN$_x$ membranes. **Eqns. (S1-2)** give the explicit equations we used to calculate the sample absorptivity

$$A^{ref} = K \cdot \frac{V_{AC}^{ref} \cdot P_{pump}^{ref}}{V_{DC}^{ref}} \tag{S1}$$

$$A^{SiN} = K \cdot \frac{V_{AC}^{SiN} \cdot P_{pump}^{SiN}}{V_{DC}^{SiN}} \tag{S2}$$

where $A^{ref}$ (known via FTIR measurements) and $A^{SiN}$ (to be determined) are the absorptivities of the reference and sample, respectively; $K$ is the correction factor; $V_{AC}^{ref}$ and $V_{AC}^{SiN}$ are AC components of measured PCI signals for the reference and the sample, respectively; $V_{DC}^{ref}$ and $V_{DC}^{SiN}$ are DC components of measured PCI signals for the reference and the sample, respectively; $P_{pump}^{ref}$ and $P_{pump}^{SiN}$ are pump-laser powers used in the PCI measurements (and measured with a power meter) for the reference and the sample, respectively. To convey an idea of the numbers involved, we show in **Table S4** data from a single position on the $Si_3N_4$ (dataset 2) and a single position on the $SiN_x$ membrane (out of the 2601 points measured on each sample).

**Table S4.** Numerical values used to determine the absorptivity of one point each on the $Si_3N_4$ and $SiN_x$ membranes

| $Si_3N_4$ (dataset 2) | | $SiN_x$ | |
|---|---|---|---|
| $A^{ref}$ | 1.5% | $A^{ref}$ | 2.6%; |
| $V_{AC}^{ref}$ | 1.65E-4 | $V_{AC}^{ref}$ | 0.03756 |
| $V_{DC}^{ref}$ | 1.375 | $V_{DC}^{ref}$ | 1.231 |
| $P_{pump}^{ref}$ | 45.5 $\mu$W | $P_{pump}^{ref}$ | 20 mW |
| $A^{sample}$ | 3.38E-7 | $A^{sample}$ | 1.94E-3 |
| $V_{AC}^{sample}$ | 1.3E-4 | $V_{AC}^{sample}$ | 0.0371 |
| $V_{DC}^{sample}$ | 1.4763 | $V_{DC}^{sample}$ | 1.248 |
| $P_{pump}^{SiN}$ | 2 W | $P_{pump}^{SiN}$ | 253 mW |
| $K$ | 5.69E-3 | $K$ | 1.70E-2 |

To determine the absorptivity of $Si_3N_4$ (dataset 2) and $SiN_x$ membranes that we report in the main text, we conducted 2601 PCI measurements on both membranes over areas of 0.5 mm × 0.5 mm. Throughout those measurements, we kept $P_{pump}^{sample}$ unchanged, and assumed $K$ did not change with respect to different scanning locations for each type of membrane. We also found the variation of $V_{DC}^{sample}$ to be negligible across different measurements. We recorded $V_{AC}^{sample}$ for each measurement and used **Eqn. (S2)** to calculate the absorptivity for each scanning location. The results of the 2D scans are shown in **Fig. 3** in the main text, and histograms of those measured absorptivities for both $Si_3N_4$ (dataset 2) and $SiN_x$ membranes are shown in **Fig. S7**.

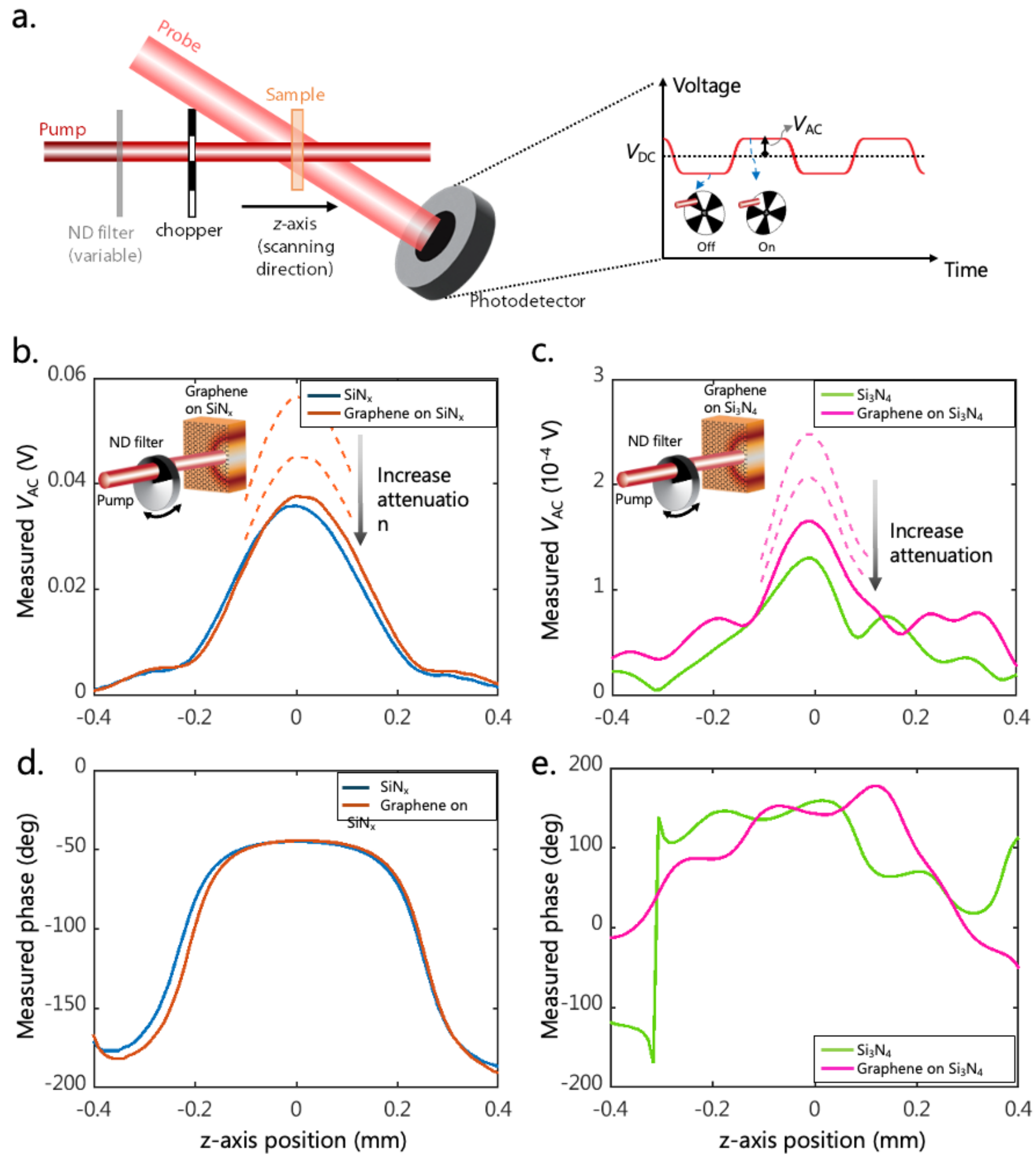

**Fig. S6.** Modified version of Fig. 2 from main text where panels (c) and (d) are using dataset 2 (rather than dataset 1, as in the main text). (a) Side-view schematic of the PCI setup showing translation of sample along the z-axis, and the AC and DC components of the detected signal. The sample is translated in the z-direction to find the peak of the AC signal which occurs when the pump waist is at the sample surface; (b) AC component of the detected probe intensity ($V_{AC}$) for the $SiN_x$ membrane with and without graphene. The pump intensity was attenuated using a variable ND-filter for the sample with graphene to obtain a similar $V_{AC}$ to that of $SiN_x$ alone (inset). Solid lines are the measured $V_{AC}$, while dashed lines represent the process of increasing attenuation to achieve similar $V_{AC}$ with and without graphene; (c) $V_{AC}$ for the $Si_3N_4$ (dataset 2) membrane with and without graphene, similarly obtained by attenuation using a variable ND-filter; (d, e) Phase between the chopped pump and detected probe intensities vs. the sample position for (d) the $SiN_x$ membrane and (e) $Si_3N_4$ membrane, with and without graphene.

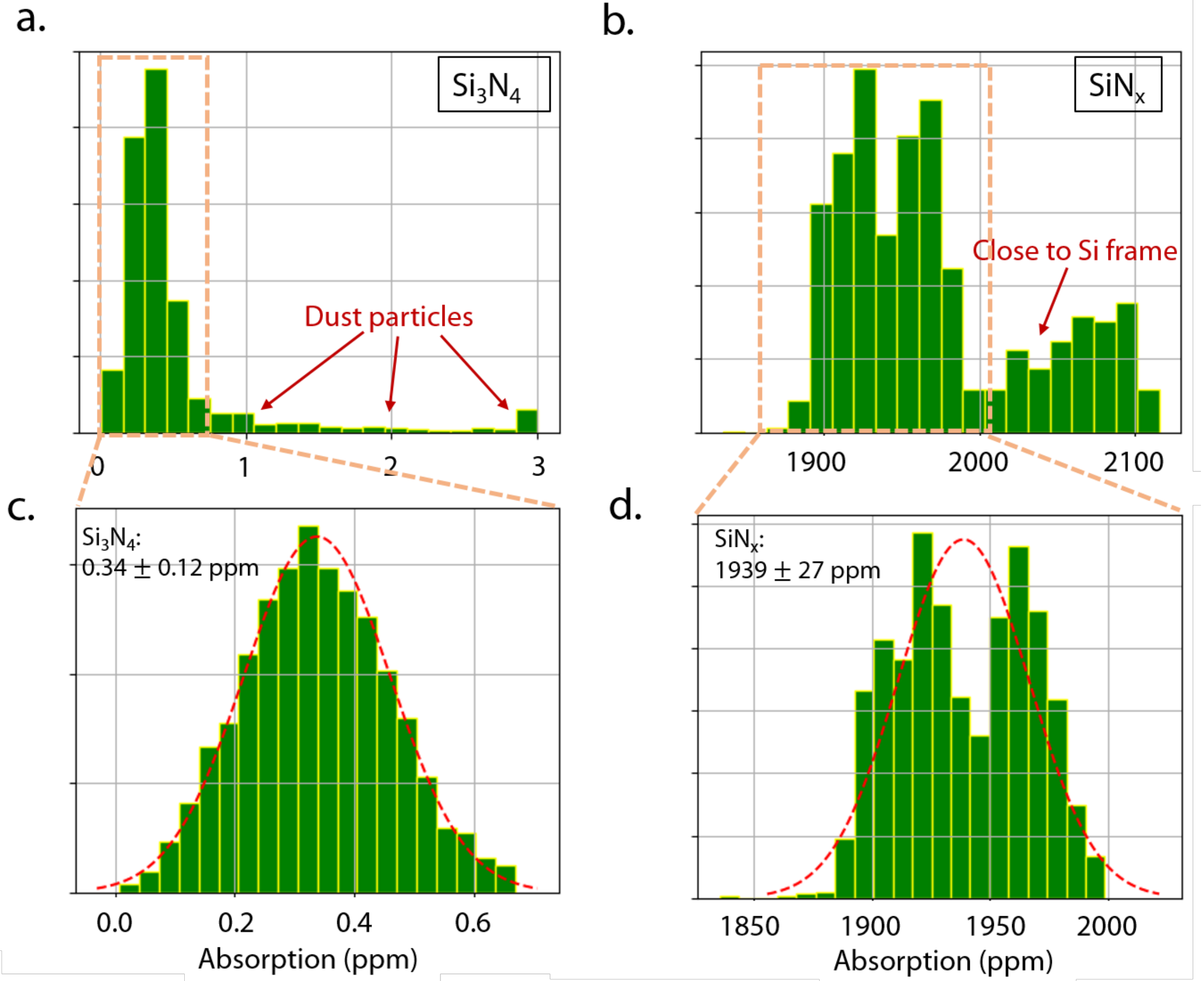

**Fig. S7**. (a-b) Histograms of 2D scans of absorptivity for (a) $Si_3N_4$ (dataset 2), and (b) $SiN_x$ ($x \sim 1$) membranes, in parts per million (ppm). (c) Histogram and a Gaussian fit of the absorptivity of $Si_3N_4$ (dataset 2), after removing data points corresponding to dust particles. (d) Histogram and a Gaussian fit of the absorptivity of $SiN_x$ ($x \sim 1$), after removing data points that are close to Si frame. In these figures, we assume $A^{ref}$ = 2.6 % for graphene-on-$SiN_x$ reference, and $A^{ref}$ = 1.5 % for graphene-on-$Si_3N_4$ reference.

As we discussed in the manuscript, for the $Si_3N_4$ (dataset 2), dust particles lead to absorption peaks in PCI measurements, and in **Fig. S7a** they correspond to the long-tail feature in the histogram. To remove those points, we discarded all data points with absorptivity > 0.67 ppm, and did a Gaussian fit on the remaining data points (**Fig. S7c**). This gives us an absorptivity of $(3.4 \pm 1.2) \times 10^{-7}$ for the $Si_3N_4$ (dataset 2). For the $SiN_x$ membrane, we observed an increase in absorptivity when the pump laser beam was close to the Si frame, and this leads to a double-peak pattern in the histogram (**Fig. S7b**). To remove those points, we discarded all data points with absorptivity > 2000 ppm, and did a Gaussian fit on the remaining data points (**Fig. S7d**). This gives us an absorptivity of $(1.94 \pm 0.027) \times 10^{-3}$ for the $SiN_x$ membrane. Note that in this paragraph, the uncertainty of the absorptivity estimate comes directly from the distributions in **Fig. S7**, and may underestimate the true uncertainty, as described in the next section.

*Error-bar calculations for $Si_3N_4$ (dataset 2) and SiNx*

In **Fig. S7**, the distribution of absorptivity values is indicative of uncertainty due to variation in the material across the sample and noise in the PCI setup. However, for us to appropriately estimate the absorption coefficient, we also need to consider the uncertainty of the measured absorptivity values of reference samples ($A^{ref}$). Here, we assume these 2 sources of uncertainty are independent.

As described in the main text, the absorption of the reference (graphene-coated SiN membrane), $A^{ref}$, is $(1.5 \pm 0.11)\%$ for graphene-on-$Si_3N_4$ and $(2.6 \pm 0.16)\%$ for graphene-on-$SiN_x$. To translate the uncertainty of $A^{ref}$ to the uncertainty of absorptivity, we repeat the process in the above subsection, with $A^{ref} = (1.5 - 0.11)\%, 1.5\%, (1.5 + 0.11)\%$ for graphene-on-$Si_3N_4$ reference, and $A^{ref} = (2.6 - 0.16)\%, 2.6\%, (2.6 + 0.16)\%$ for graphene-on-$SiN_x$ reference, and calculate the standard deviations of the absorptivities of $Si_3N_4$ (dataset 2) and $SiN_x$ with those different $A^{ref}$ values (denoted as $\sigma_A^{Si_3N_4}$ and $\sigma_A^{SiN_x}$, respectively).

The total error bar of absorptivity, $\sigma_{total}^{Si_3N_4}$ and $\sigma_{total}^{SiN_x}$, can be estimated using equations $\sigma_{total}^{Si_3N_4} = \sqrt{\left(\sigma_A^{Si_3N_4}\right)^2 + \left(\sigma_{PCI}^{Si_3N_4}\right)^2}$, $\sigma_{total}^{SiN_x} = \sqrt{\left(\sigma_A^{SiN_x}\right)^2 + \left(\sigma_{PCI}^{SiN_x}\right)^2}$, where $\sigma_{PCI}^{Si_3N_4}$ and $\sigma_{PCI}^{SiN_x}$ are standard deviations from PCI measurements calculated in the previous subsection ($1.2 \times 10^{-7}$ for $Si_3N_4$, and $0.027 \times 10^{-3}$ for $SiN_x$). Using this approach, we obtained the absorptivity of $Si_3N_4$ (dataset 2) to be $(3.4 \pm 1.23) \times 10^{-7}$, and the absorptivity of $SiN_x$ membrane to be $(1.94 \pm 0.12) \times 10^{-3}$.

*Absorption-coefficient calculations for $Si_3N_4$ (dataset 2) and SiNx*

To convert the absorptivities of the membranes to the absorption coefficients of $Si_3N_4$ (dataset 2) and $SiN_x$, we applied the transfer-matrix method[S12] on an infinitely wide single-layer membrane surrounded by air as illustrated in **Fig. S8**. For the $Si_3N_4$ (dataset 2), the thickness used was 194.4 nm (fitted from ellipsometry, **Table S2**), and the complex refractive index at 1064 nm is $2.004 + \frac{\alpha_{Si_3N_4} \cdot \lambda}{4\pi} \cdot i$, where $\lambda = 1064\ nm$, and $\alpha_{Si_3N_4}$ is the unknown to be solved. Using the root finder, we found when $\alpha_{Si_3N_4} = (2.09 \pm 0.76) \times 10^{-2}$ $cm^{-1}$, the membrane had an absorptivity of $(3.4 \pm 1.23) \times 10^{-7}$ (corresponding to the PCI measurement result). Similarly, for the $SiN_x$ membrane, the thickness used was 2.01 $\mu$m, and the complex refractive index used at 1064 nm was $2.147 + \frac{\alpha_{SiN_x} \cdot \lambda}{4\pi} \cdot i$, where $\lambda = 1064\ nm$, and $\alpha_{SiN_x}$ was the unknown to be solved. Using the root finder, we found when $\alpha_{SiN_x} = 7.94 \pm 0.50$ $cm^{-1}$, the absorptivity of the membrane was the same as our PCI measurement result ($(1.94 \pm 0.12) \times 10^{-3}$).

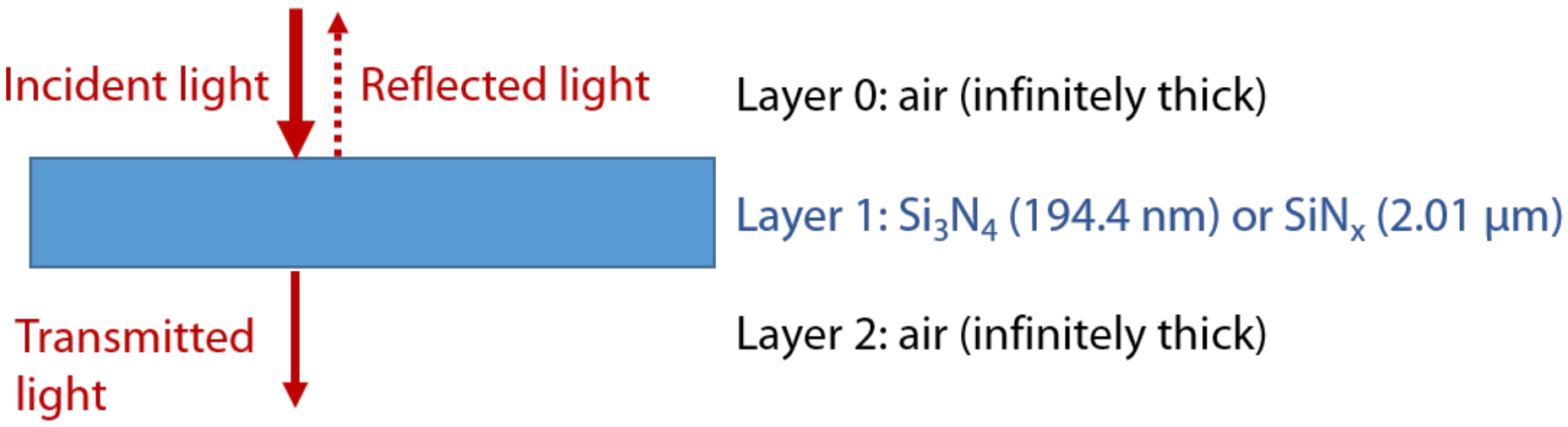

**Fig. S8**. Illustration of the model used in the transfer-matrix calculation to determine the absorption coefficients of $Si_3N_4$ and $SiN_x$.

*Response of $V_{AC}$ and phase to pump power*

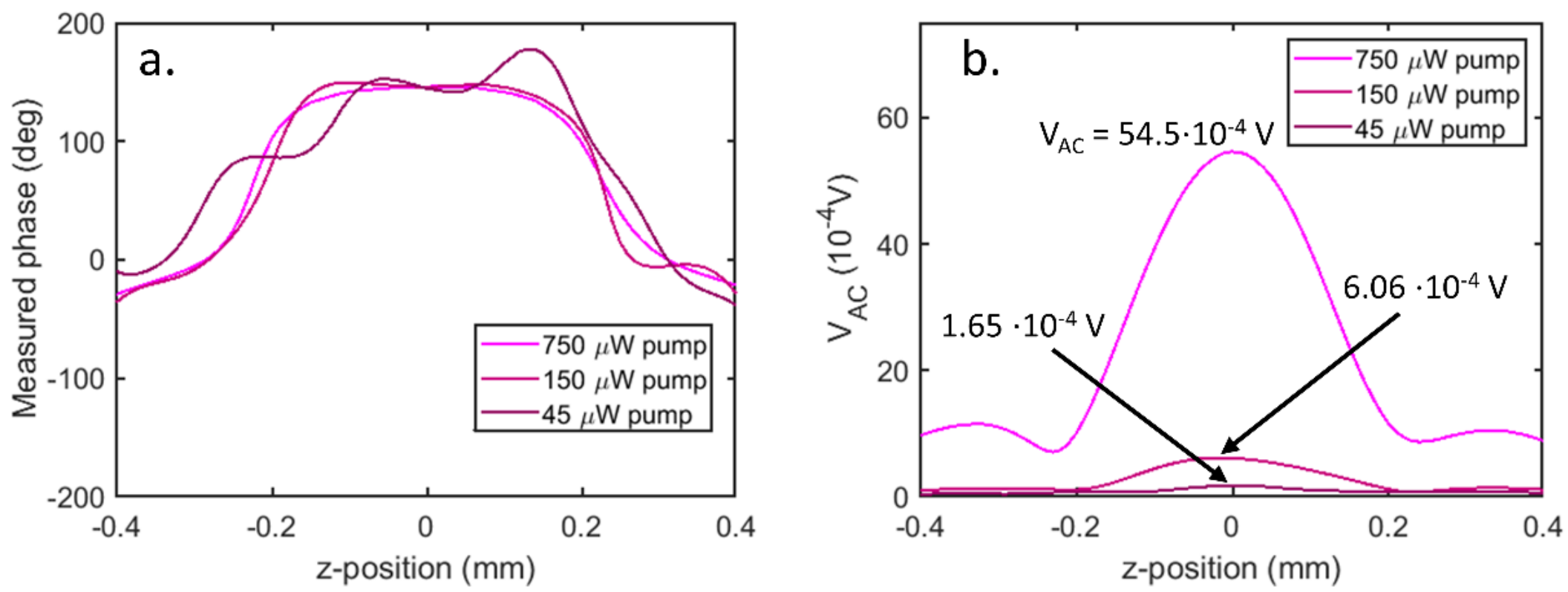

**Fig. S9** (a) Phase vs z-position plot for the PCI scan of $Si_3N_4$ coated with monolayer graphene (dataset 2) at different pump powers. The phase is independent of the pump power, although the trend becomes noisier for low pump powers (45 μW in this case) due to low signal-to-noise ratio. (b) $V_{AC}$ for different pump powers. In our self-referencing PCI experiment, the pump was manually attenuated to ~45 μW for $Si_3N_4$-with-graphene to achieve a $V_{AC}$ close to that of bare $Si_3N_4$ with pump power of 2 W. We note that **Fig. S9** is based on data from dataset 2 of graphene-coated $Si_3N_4$, which is why the $V_{AC}$ and phase numbers are different from what is in the current version of the main text in **Fig. 2**—where dataset 1 is used.

**Fig. S9** shows how phase and $V_{AC}$ change as the pump power is increased more than an order of magnitude from ~45 μW to ~750 μW for graphene on $Si_3N_4$. **Fig. S9(a)** shows that the phase is largely independent of pump power, with the signal becoming noisier at the lowest pump power of ~45 μW. **Fig. S9(b)** shows how the $V_{AC}$ to pump-power ratio changes for a range of powers across more than an order of magnitude. **Table S5** below summarizes data from **Fig. S9(b)**:

**Table S5.** $V_{AC}$ and $V_{AC}$ to pump-power ratio for a graphene on $Si_3N_4$ sample

| Pump power (μW) | $V_{AC}$ (10$^{-4}$V) | $V_{AC}$ to pump ratio (10$^{-6}$V/μW) |
|---|---|---|
| 45 | 1.65 | 3.7 |
| 150 | 6.06 | 4 |
| 750 | 54.5 | 7.3 |

We note that the $V_{AC}$ to pump power ratio increases and therefore the photothermal effect as a function of pump is non-linear as pump power is increased more than an order of magnitude (shown for graphene on $Si_3N_4$ in **Table S5**). However, by design this is not a problem in self-referencing PCI, because we conduct experiments at similar $V_{AC}$ between the reference sample and the sample being tested (graphene-$Si_3N_4$ and $Si_3N_4$ in this work), which corresponds to similar amounts of photothermal lensing. In this way, self-referencing PCI ensures that we always operate in quasi-linear regimes.

## S6. Comparison of measured loss in $Si_3N_4$ with other works

**Table S6.** Loss and absorption coefficient of $Si_3N_4$ reported in other papers

| **Work** | **Method** | **Growth** | **Wavelength (nm)** | **Loss in dB cm$^{-1}$** | **Abs. coefficient (cm$^{-1}$)** |
|---|---|---|---|---|---|
| Our work | Self-referencing PCI | LPCVD (purchased from Norcada Inc.) | **1064** | $5.43\times10^{-2}$ | $\sim1.25\times10^{-2}$ |
| Land/Wilson 2024[S13] | Nanomechanical absorption spectroscopy | LPCVD | **532, 633, 750, 780, 800, 850, 1064, 1550** | 1.36, 0.63, 0.34, 0.29, 0.26, 0.23, 0.09, 0.24 | 0.31, 0.15, 0.08, 0.07, 0.06, 0.05, 0.02, 0.06 |
| Ikeda/ Fainman 2008[S14] | Measure loss vs different waveguide lengths | PECVD | **1548** | 4 (transmission loss) | 0.92 |
| Ji/ Lipson 2017[S15] | Cavity ringdown; compare losses in different structures | LPCVD | **1560** | $(1.3\pm0.5)\times10^{-3}$ (bulk) | $\sim3\times10^{-4}$ |
| Luke/ Lipson 2013[S16] | $\alpha$ calculated from Q factor | LPCVD | **1550** | $4.2\times10^{-2}$ (transmission loss); $2.94\times10^{-2}$ (absorption loss) | $9.7\times10^{-3}$ (transmission loss); $6.8\times10^{-3}$ (absorption loss) |

## S7. Application in light sails

Here we calculate the equilibrium temperature of a hypothetical light sail under laser illumination, modeled as an infinitely-wide thin membrane (**Fig. S10(a)**). The membrane thicknesses of such hypothetical light-sails are chosen such that the reflectivities of the sails are maximized. Using the transfer-matrix method, we determined the optimal thickness of $Si_3N_4$-based light sail to be 132.68 nm with a reflectivity of 0.36, and the optimal thickness of $SiN_x$-based light sail to be 123.57 nm with a reflectivity of 0.42.

In a sail under illumination, two major thermal processes happen. First, the sail absorbs incident light due to its non-zero absorption coefficient, leading to an increase in the temperature. Second, the sail emits electromagnetic waves that lead to cooling (known as radiative cooling) (**Fig. S10(a)**). The equilibrium temperature $T_{eq}$ is reached when the power absorbed is equal to the power emitted and, to maintain sail integrity, the upper limit of $T_{eq}$ should be lower than the ultra-high vacuum (UHV) melting temperature of $Si_3N_4$[S17] Here we set the upper limit of $T_{eq}$ to be 1500 K, which is on the lower end of the estimates of the decomposition temperature of $Si_3N_4$[S18,S19].

We thus calculated the equilibrium temperature $T_{eq}$ reached by each sail using Eqn. S3:

$$A \cdot P_{inc} = 2A_{sail} \int_a^b \frac{c_1}{\lambda^5} \cdot \frac{\epsilon_{sail}(\lambda)}{\exp\left(\frac{c_2}{\lambda \cdot T_{eq}} - 1\right)} d\lambda \quad \text{( S3 ).}$$

The left-hand side is power absorbed by the sail ($P_{abs}$), and the right-hand side is emitted black-body radiation ($P_{em}$), where $A$ is the absorbance of the sail, $P_{inc}$ is the incident laser power on the sail, $A_{sail}$ is the area of a single side of the sail, the factor of 2 accounts for the emission on both sides of the sail, $a = 1.4\ \mu m$ and $b = 32\ \mu m$ are the integration bounds that capture the vast majority of thermal radiation and are the bounds for SiN material properties from Luke et al.[S20], $\epsilon_{sail}(\lambda)$ is the wavelength-dependent spectral emissivity of the sail, $c_1 = 2\pi hc^2$, $c_2 = hc/k_b$, $h$ is Planck's constant, $c$ is the speed of light, and $k_b$ is the Boltzmann's constant. Here, $A$ for a given sail geometry is calculated using the transfer-matrix method and the absorption coefficient measured by PCI. $\epsilon_{sail}(\lambda)$ for a given sail is calculated using the transfer-matrix method and the complex refractive index of $Si_3N_4$[S12]. For simplicity, we assume $\epsilon_{sail}(\lambda)$ for $SiN_x$ to be the same as that for $Si_3N_4$.

The absorbed and emitted power densities under 10 GW/m$^2$ of illumination (which is the driving laser power density in the Breakthrough Starshot mission)[S21] as a function of temperature are plotted in **Fig. S10(b)** for $Si_3N_4$, and in **Fig. S10(c)** for $SiN_x$. The temperature at which $P_{abs} = P_{em}$ is the equilibrium temperature $T_{eq}$. For our highest reported $Si_3N_4$ absorbance of $3\times10^{-2}$ cm$^{-1}$, we observed the equilibrium temperature $T_{eq}$ of the $Si_3N_4$ sail to be approximately 940 K, lower than the decomposition temperature (1500 K) of $Si_3N_4$. It must be noted however that the calculated equilibrium temperature will likely change with the use of temperature-dependent absorption data and more optimized sail designs[S21,S22]. We note that for $SiN_x$, an equilibrium temperature is not achieved at the GW/m$^2$ power scale. We also calculated the equilibrium temperature as a function of incident laser intensity. To achieve a similar equilibrium temperature for $SiN_x$ as for $Si_3N_4$, the laser power must be ~3 orders of magnitude lower (**Fig. S10(d)**).

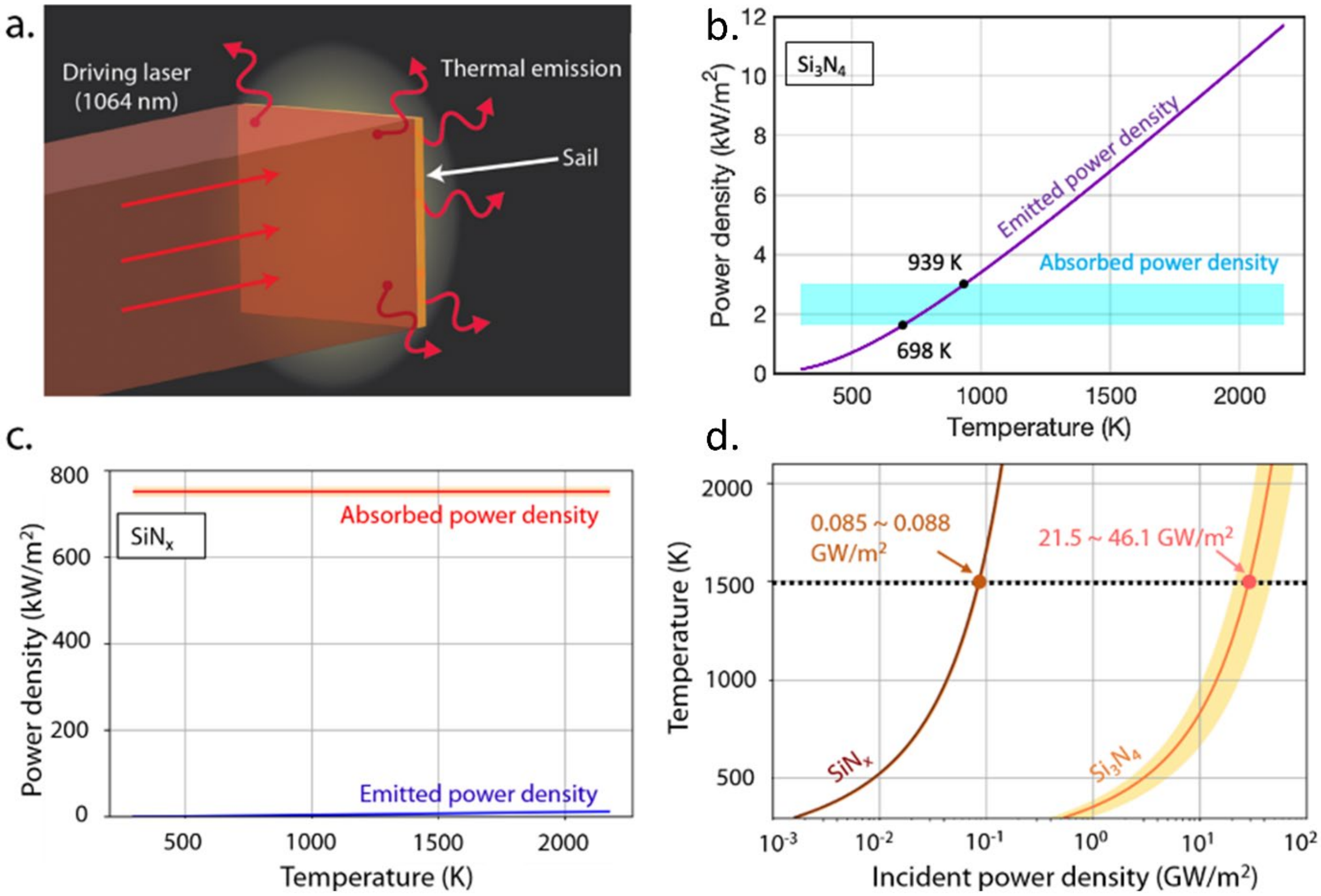

**Fig. S10**. (a) Schematic of a silicon-nitride sail in space illuminated by a 1064 nm laser used in our thermal-equilibrium calculation. In our calculation, both the silicon-nitride sail and the incident laser are infinitely wide, and the incident laser has a uniform power density. (b) Absorbed and emitted power density vs sail temperature for a $Si_3N_4$ sail, illuminated by a 10 GW/m$^2$ laser with $\lambda = 1064$ nm. The shaded cyan region represents absorbed power density for the absorptivity range of $Si_3N_4$ presented in the main text. The marked temperatures are points of equilibrium for lowest and highest calculated absorptivities, where absorbed and emitted power densities are equal. (c) Absorbed and emitted power density vs sail temperature for a $SiN_x$ sail, illuminated by a 10 GW/m$^2$ laser with $\lambda = 1064$ nm. The orange shaded region corresponds to the error bar of the absorbed power density, due to the uncertainty of absorption coefficient of $SiN_x$. No thermal equilibrium can be reached within the 300-2100 K temperature range for this $SiN_x$ sail. (d) Equilibrium temperature as a function of incident laser intensity for the $Si_3N_4$ sail and the $SiN_x$ sail considered in (b) and (c). The dashed line corresponds to the decomposition temperature of $Si_3N_4$ at 1500 K. Shaded regions correspond to error bars of incident laser intensity, due to the uncertainty of absorption coefficient of $Si_3N_4$ and $SiN_x$.

## S8. Supplementary References